\documentclass[aip,jcp,amsmath,amssymb,reprint ]{revtex4-2}
\usepackage{mathtools}
\usepackage{hyperref}
\usepackage{graphicx}
\usepackage{listings}
\usepackage{color}

\usepackage[caption=false]{subfig}
\definecolor{dkgreen}{rgb}{0,0.6,0}
\definecolor{gray}{rgb}{0.5,0.5,0.5}
\definecolor{mauve}{rgb}{0.58,0,0.82}

\begin{document}

\title{Bayesian Sampling of Structural Ensembles: The Role of Ensemble-Counting Measures}

\author{Ivan Gilardoni}
\affiliation{Scuola Internazionale Superiore di Studi Avanzati, Via Bonomea 265, 34136 Trieste, Italy}
\author{Giovanni Bussi}
\email{bussi@sissa.it}
\affiliation{Scuola Internazionale Superiore di Studi Avanzati, Via Bonomea 265, 34136 Trieste, Italy}

\begin{abstract}
Structural ensemble refinement is widely used to integrate molecular simulations with experimental measurements. While most applications focus on the maximum-a-posteriori (MAP) ensemble, Bayesian sampling of the posterior distribution can provide uncertainty estimates and posterior averages for arbitrary observables. A notable step in this direction was introduced by the Bayesian Energy Landscape Tilting (BELT) framework, where sampling is performed on a family of maximum-entropy ensembles parametrized by Lagrange multipliers.
Here, we show that Bayesian sampling in this setting requires an explicit choice of ensemble-counting measure. In particular, the flat measure in Lagrange-multiplier space used in the original BELT formulation leads to a posterior distribution that is formally non-normalizable for finite reference trajectories. We propose the Jeffreys measure as an invariant ensemble-counting prescription, restoring normalizability in the finite-sample situations considered here, and providing a consistent definition of posterior averages. Using both an analytically tractable Gaussian model and maximum-entropy refinement of RNA oligomer simulations, we compare different ensemble-counting measures and show that they can significantly affect Bayesian estimates. The resulting methodology has been implemented in the \texttt{MDRefine} software package.
\end{abstract}

\maketitle

\section{Introduction}

Molecular dynamics (MD) simulations are an essential tool for investigating the structural and dynamical properties of biomolecular systems. Their predictive power, however, is limited by both the precision of the generated ensembles and the accuracy of the underlying force fields. The former can be improved through enhanced-sampling techniques~\cite{henin2022enhanced}, whereas the latter can be addressed by integrating simulations with experimental data~\cite{bottaro2018biophysical}. In this context, ensemble-refinement and reweighting approaches have become important tools in integrative structural biology, enabling the combination of prior information from molecular simulations with heterogeneous experimental measurements~\cite{rozycki2011saxs,pitera2012use,hummer2015bayesian,bottaro2018conformational,kofinger2019efficient,bottaro2020integrating,bottaro2020integratingnar,reisser2020conformational,pesce2021refining,bernetti2021reweighting,stelzl2022global,bergonzo2022conformational,frohlking2023simultaneous,tang2023ensemble,oxenfarth2023integrated,kofinger2024encoding,borthakur2025determining,muller2025unique,raddi2025model,silva2026cryo}.
In ensemble refinement, one seeks a refined ensemble \(P\) that improves agreement with experimental data while remaining close to a reference ensemble \(P_0\). This is commonly formulated by minimizing a loss function composed of two terms: a term measuring discrepancy with experiment, often a \(\chi^2\), and a regularization term given by the relative entropy between \(P\) and \(P_0\) \cite{hummer2015bayesian}. The balance between these two contributions is controlled by a hyperparameter. This formulation is closely related to the maximum-entropy principle: among the ensembles compatible with the experimental information, one selects the least informative modification of the reference ensemble~\cite{cesari2018using}.
There are different ways to represent the refined ensemble.
One possibility is to represent it through a set of weights assigned to the configurations of a reference ensemble.
This strategy has been employed in several contexts, including SAXS~\cite{rozycki2011saxs} and, more recently, cryo-EM ensemble reweighting~\cite{tang2023ensemble,silva2026cryo}. Sampling or optimizing the weights provides a highly general representation of refined ensembles.
However, most practical maximum-entropy reweighting approaches exploit an equivalent parametrization in terms of Lagrange multipliers associated to the restrained observables \cite{pitera2012use,cesari2018using,bottaro2020integrating}. This reduces the dimensionality of the optimization from the number of trajectory frames to the number of experimental observables. For maximum-a-posteriori refinement, this parametrization is usually advantageous and does not change the optimized ensemble.

The same loss function can also be interpreted within Bayesian inference. In this view, after specifying a reference measure on the space of ensembles, the exponential of minus the loss is proportional to a posterior density given the reference simulation and the experimental data. With the conventional reference measure implicit in the original formulation, minimizing the loss gives the corresponding maximum-a-posteriori (MAP) ensemble. However, the MAP estimate provides only a single refined ensemble. A full Bayesian treatment would instead sample the posterior distribution over possible refined ensembles, thereby providing posterior averages and uncertainty estimates for arbitrary observables chosen after refinement.
A notable step in this direction was introduced by the Bayesian Energy Landscape Tilting (BELT) framework~\cite{beauchamp2014bayesian}. BELT restricts posterior sampling to the maximum-entropy family
\(P_{\vec\lambda}\) parametrized by the Lagrange multipliers (or generalized forces) \(\vec\lambda\).
This strategy is often computationally advantageous because it replaces sampling in the high-dimensional space of ensemble weights with sampling over the usually smaller set of Lagrange multipliers.
However, once posterior averages rather than only MAP estimates are considered, an additional issue appears: one must specify how ensembles are counted on the sampled manifold.
Different measures in  \(\vec\lambda\) can lead to different Bayesian averages and variances.

Here we show that this ensemble-counting issue has concrete consequences. In particular, for a finite reference trajectory, the flat-\(\vec\lambda\) posterior used in the original BELT formulation is formally not normalizable. At large values of \(|\vec\lambda|\), the refined ensemble collapses onto one or a few configurations; the ensemble then changes negligibly as \(\vec\lambda\) is further increased, while the posterior density remains finite over an unbounded region of parameter space. Thus, the same collapsed ensemble is effectively counted many times. This does not affect the MAP solution, but it makes posterior averages and uncertainties ill defined unless a suitable ensemble-counting measure is specified.
We propose to use the Jeffreys prior \cite{jeffreys1946invariant} as an invariant ensemble-counting measure on the restricted manifold
$P_{\vec\lambda}$.
The Jeffreys measure is induced by the Fisher information metric, or equivalently by the local distinguishability between nearby probability distributions. It therefore assigns posterior volume according to changes in the ensemble itself rather than according to an arbitrary parametrization. This restores normalizability of the posterior in the finite-sample cases considered here and makes posterior averages invariant under reparametrization of the same ensemble manifold.
We analyze the impact of ensemble counting explicitly. First, we use a Gaussian model, where the origin of the non-normalizability and its dependence on the number of frames can be understood analytically. We then test the different ensemble-counting prescriptions on a realistic maximum-entropy refinement of RNA oligomer simulations using experimental scalar couplings.
Posterior sampling with these measures is implemented
in the \texttt{MDRefine} package~\cite{gilardoni2025mdrefine}.

\section{Methods}

\subsection{Bayesian-inference of ensembles and maximum a-posteriori}

In the framework of ensemble refinement, a common approach consists of minimizing
a loss function that can be introduced following the Bayesian inference of ensemble (BioEn) method \cite{hummer2015bayesian}:
\begin{equation}
\mathcal L[P] = \frac{1}{2}\chi^2[P] - \alpha S[P||P_0]
\label{eqn:loss_ER}
\end{equation}
Here $\chi^2$ quantifies the agreement between the ensemble averages $\langle g_j(x)\rangle_P$ and the corresponding experimental data $g_{j,exp}\pm\sigma_{j,exp}$, hence has a functional dependence on $P$. 
The relative entropy $S[P||P_0]$ enforces the closeness between $P$ and the reference ensemble $P_0$. A hyperparameter $\alpha>0$ tunes the relative relevance of the two terms, ranging between complete trust in the experiment ($\alpha \to 0$) or in the prior distribution ($\alpha\to+\infty$).
In fact, this expression is related to the regularized maximum entropy method \cite{cesari2018using,frohlking2023simultaneous}: for any positive $\alpha$, among all the ensembles equally in agreement with experiment, the one with the largest relative entropy \cite{caticha2004relative} is chosen.
Given this, it is possible to show that the minimum of $\mathcal{L}$ has the following functional form:

\begin{equation}
P_\lambda(x) = \frac{1}{Z_\lambda} P_0(x) e^{-\vec\lambda\cdot\vec g(x)}
\label{eqn:p_lambda}
\end{equation}
where the coefficients $\lambda_j$ are defined by the set of implicit equations:

\begin{equation}
\lambda_j = \frac{\langle g_j \rangle_{P_\lambda} - g_{j,exp}}{\alpha\sigma_{j,exp}^2}
\label{eqn:lambda_impl}
\end{equation}
and $Z_\lambda$ is the normalization factor. This parametrization can be exploited when minimizing
$\mathcal{L}$. Indeed, rather than exploring all the possible functional forms of $P$, the optimal ensemble can be determined by minimizing over the $\lambda$ parameters. The coefficients $\lambda_j$ are the generalized forces introduced by Hummer and K\"ofinger~\cite{hummer2015bayesian} and subsequently used as optimization variables by K\"ofinger et al.~\cite{kofinger2019efficient}; equivalently, they are the Lagrange multipliers conjugate to the restrained observables in the maximum-entropy formulation with experimental errors~\cite{pitera2012use,cesari2016combining}. This parametrization effectively reduces the dimensionality of the problem.

\subsection{Full Bayesian sampling}
In the context of Bayesian inference, the loss function in Eq.~\ref{eqn:loss_ER} can be interpreted as a negative log-posterior density \cite{hummer2015bayesian}. Once a reference measure on the space of ensembles is specified, $e^{-\mathcal L[P]}$ is proportional to the posterior density $\mathcal P$ of an ensemble $P$ given the initial assumption $P_0$ and the experimental data:
\begin{equation}
\mathcal{P}(P)=\frac{1}{\mathcal Z}\,e^{-\mathcal L[P]}
\end{equation}
Here, $\mathcal{P}(P)$ is a density with respect to the measure $d\mu(P)$ on the space of ensembles, and $\mathcal Z = \int d\mu(P)\,e^{-\mathcal L[P]}$ is its normalization factor. In a continuous ensemble space, an individual ensemble generally has zero posterior probability; probabilities are assigned to regions, while the MAP refers to the maximum of a density relative to a specified measure. Throughout this work, we retain the conventional definition of the MAP ensemble as the minimizer of the original loss $\mathcal L[P]$; the alternative ensemble-counting measures introduced below are used to define posterior averages and uncertainties.
Rather than limiting to the MAP estimate, one could be interested in sampling the entire posterior distribution $\mathcal P(P)$ over all the possible ensembles $P$, in order to get the Bayesian expected value and uncertainty for a given observable. Formally, this requires computing expectation values in the form

\begin{equation}
\begin{split}
\mathbb E[\langle O \rangle_P]
&= \int d\mu(P)\,\mathcal P(P) \langle O \rangle_P \\
\end{split}
\label{eqn:bayes_av}
\end{equation}
For an explicitly discrete set of candidate ensembles, the corresponding expectation would instead be a sum over their posterior probability masses. In the setting considered here, however, the ensembles vary continuously. If $P$ is a probability distribution over a continuous configuration space, this expression represents a functional integral. For a finite reference trajectory, $P$ is represented by a finite vector of continuously varying weights and the integral is over the corresponding probability simplex. In the BELT framework considered below, the integration is further restricted to the finite-dimensional maximum-entropy family $P_{\vec\lambda}$ and becomes an ordinary integral over $\vec\lambda$. In every case, the integral is defined only after specifying the measure $d\mu(P)$. This choice is an essential part of the Bayesian sampling procedure and constitutes the ensemble-counting problem investigated here. In the following, we compare different choices of $d\mu(P)$, which specify how volume is counted on the ensemble manifold. Relative to a fixed coordinate measure, such as $d\vec\lambda$, this choice can equivalently be expressed as a density factor and is therefore often called a ``prior.'' We use the term ``ensemble-counting measure'' to emphasize its geometric role and to distinguish it from the factor $e^{\alpha S[P||P_0]}$, which encodes the prior preference for ensembles close to the reference ensemble $P_0$. Thus, this distinction concerns the roles assigned to the two factors; mathematically, both contribute to the prior probability measure over ensembles.

\subsection{Bayesian energy landscape tilting (BELT) measure}
A computationally efficient strategy for Bayesian posterior sampling was introduced in the Bayesian Energy Landscape Tilting (BELT) framework.\cite{beauchamp2014bayesian}
Rather than sampling arbitrary ensemble weights, BELT restricts the posterior to the family \(P_{\lambda}\) of maximum-entropy ensembles parametrized by the Lagrange multipliers
$\vec{\lambda}$ as described in Eq. \ref{eqn:p_lambda}.
This is equivalent to say that among all the ensembles with a set of 
average values $\langle\vec g \rangle_P$ of the observables,
only the one maximizing the relative entropy from the reference ensemble $P_0$ will be considered.
Then, they propose to sample the posterior $\mathcal P(\vec\lambda)\propto e^{-\mathcal L(P_\lambda)}$ with a uniform measure with respect to $\vec\lambda$

\begin{equation}
\mathbb E[\langle O \rangle_P]_{\textrm{BELT}}
= \frac{1}{\mathcal Z} \int d\vec\lambda\, e^{-\mathcal{L}[P_{\vec\lambda}]} \langle O \rangle_{P_{\vec\lambda}}.
\label{eqn:bayes_av_belt}
\end{equation}

While this approach may appear straightforward, it actually reflects a specific choice for how the posterior averages are calculated: namely, by interpreting $e^{-\mathcal L(P_\lambda)}$ as the probability density of ensembles within an infinitesimal volume $d\vec\lambda$ around $\vec\lambda$ (up to a constant prefactor).
At very high values of $|\vec\lambda|$, the loss function, and thus the posterior, does not change significantly as a function of $\vec\lambda$, reaching an asymptotic plateau value, rather than diverging. Consequently, the posterior is not normalizable because the ensemble shows minimal variation across these regions, and so one basically ends up repeatedly sampling the same ensemble multiple times. This will become particularly evident in the simple Gaussian model discussed below.

\subsection{Probability of reaching the plateau in a Gaussian model}
\label{sec:escaping}

Whereas having a non-normalizable posterior makes Bayesian sampling not well defined,
it is still interesting to consider how likely a Monte Carlo sampling on the ensembles would 
result in reaching the region where the probability is flat.
This can be done in a quasi-analytical way on a Gaussian model, where it is possible to compute the minimum and the asymptotic values of the loss function.
We consider as a reference ensemble the one generated by a quadratic potential:
\begin{equation}
V(x) = \frac{x^2}{2\sigma^2}
\end{equation}
where we enforce $\langle x \rangle = x_{exp} \pm \sigma_{exp}$ (see Section~\ref{sec:quadratic} for further details). The minimum of the loss function $\mathcal{L}$ can be obtained solving Eq. \ref{eqn:lambda_impl}. For a quadratic potential, assuming a sufficiently large number of frames,
this is well approximated by the continuum-limit solution:
\begin{equation}
\lambda^* = -\frac{x_{exp}}{\sigma^2 + \alpha\sigma_{exp}^2}
\label{eqn:lambda_star}
\end{equation}
which in turn results in the following ensemble average:
\begin{equation}
\langle x \rangle_{\lambda^*}=\frac{x_{exp}\sigma^2}{\sigma^2 + \alpha\sigma_{exp}^2}
\end{equation}
and in the following value for the $\chi^2$:
\begin{equation}
\chi^2=
\frac{\left(
\langle x \rangle_{\lambda^*} - x_{exp}
\right)^2}{\sigma_{exp}^2}=
\left(
\frac{x_{exp}\alpha\sigma_{exp}}{\sigma^2 + \alpha\sigma_{exp}^2}
\right)^2
\end{equation}
The Kullback-Leibler divergence can also be computed analytically.
For two Gaussian distributions with the same variance $\sigma^2$ and difference between their means $\Delta \mu$, one has $D_{\mathrm{KL}}=\frac{(\Delta \mu)^2}{2\sigma^2}$.
Since the relative entropy is the negative of the Kullback-Leibler divergence, one has
\begin{equation}
S[P_{\lambda^*}||P_0]=-
\frac{1}{2\sigma^2}\left( \frac{x_{exp}\sigma^2}{\sigma^2 + \alpha\sigma_{exp}^2}
\right)^2
=
-\frac{1}{2}\left( \frac{x_{exp}\sigma}{\sigma^2 + \alpha\sigma_{exp}^2} \right)^2
\end{equation}
In this limiting case, the value of the loss function $\mathcal{L}$ in its minimum can be shown to be
\begin{equation}
    \mathcal{L}(\lambda^*)=
\frac{\alpha x_{exp}^2}{2(\sigma^2 + \alpha\sigma_{exp}^2)}
\end{equation}
which is independent of the number of frames being reweighted.

The value of the loss function in the limit of large $|\lambda|$ is more complicated to estimate,
because it grows with the number of frames that are being reweighted.
We assume without loss of generality that $x_{exp}>0$ so that the loss function for large negative $\lambda$ is smaller than the loss function for large positive $\lambda$.
In the limit of infinite $\lambda$, a single frame will contribute to the refined distribution.
The asymptotic value of the average $x$ over the refined distribution is
$
    \lim_{\lambda \rightarrow -\infty}\langle x \rangle = x_{max}(N,\sigma)
$,
where $x_{max}(N,\sigma)$ is the maximum of $N$ values drawn from a normal distribution
of average 0 and variance $\sigma^2$. Its leading order in $N$ is $\sigma\sqrt{2\log N}$.
For large $N$, the $\chi^2$ contribution will be dominated by this divergent term.
At the same time, having a reweighted ensemble consisting of a single frame results in a relative entropy $S[P||P_0]=-\log N$.
Hence, for large $N$:
\begin{equation}
    \lim_{\lambda \rightarrow -\infty} \mathcal{L}(\lambda)%
    \approx \left[(\sigma/\sigma_{exp})^2 + \alpha \right] \log N
\label{eqn:scaling_losslim}
\end{equation}
Since $\mathcal L(\lambda^*)$ is independent of the number of frames $N$,
$\Delta\mathcal L = \mathcal L_{lim} - \mathcal L(\lambda^*)$ also grows logarithmically with the number of reweighted frames.
Whenever the experimental error $\sigma_{exp}$ is large enough or the $\alpha$ regularization term
is chosen small enough, or for small sample sizes, reaching the limiting value during Bayesian sampling will be very likely. Reaching the plateau of the posterior over $\lambda$ in turn
will result in repeatedly sampling almost the same ensemble several times.

\subsection{Jeffreys measure}

The flat-$\lambda$ measure used in BELT counts ensembles uniformly in the coordinates $\vec\lambda$. However, equal volumes in $\vec\lambda$ space do not generally correspond to equal variations of the underlying ensembles. In particular, when $|\vec\lambda|$ becomes large, many different values of $\vec\lambda$ correspond to nearly indistinguishable ensembles, often dominated by the same few frames.
These regions should therefore contribute little posterior volume, rather than being counted repeatedly as distinct ensembles.

A natural way to define a local ensemble-counting measure is to use the distinguishability between nearby probability distributions \cite{jeffreys1946invariant}. The Kullback-Leibler divergence \(D_{\mathrm{KL}}\) provides such a local notion of distinguishability.
For a normalized perturbation, \(\sum_x \delta P(x)=0\), the linear term in its expansion vanishes and we get
\begin{equation}
D_{\mathrm{KL}}[P+\delta P\,|\,P] = \frac{1}{2}\sum_x \frac{\delta P(x)^2}{P(x)} + O(\delta P^3).
\end{equation}

The quadratic form in this expansion defines an infinitesimal squared distance \(ds^2\) between nearby distributions. Restricting this distance to the parametrized family \(P_{\vec\lambda}\), one obtains

\begin{equation}
\begin{split}
ds^2 & = \sum_x \frac{d P(x)^2}{P(x)} \\ & = \sum_x P(x) \Bigl(\frac{d P(x)}{P(x)}\Bigr)^2 = \sum_x P(x) (d\log P(x))^2 \\
& = \sum_x P(x) \sum_{j,k} \frac{\partial\log P(x)}{\partial \lambda_j} \frac{\partial\log P(x)}{\partial \lambda_k} d\lambda_j d\lambda_k \\ & = \sum_{j,k} F_{jk}(\vec\lambda) \, d\lambda_j d\lambda_k
\label{eqn:ds2_jeff}
\end{split}
\end{equation}
where
\begin{equation}
F_{jk}(\vec\lambda) = \Bigl\langle \frac{\partial \log P(x)}{\partial \lambda_j} \frac{\partial \log P(x)}{\partial \lambda_k} \Bigr\rangle_{P_\lambda}
\label{eqn:der_Fisher}
\end{equation}
is the Fisher information matrix \cite{jeffreys1946invariant,jaynes1968prior}. 
For Eq.~\ref{eqn:p_lambda}, \(\partial_{\lambda_j}\log P_\lambda(x)=-(g_j(x)-\langle g_j\rangle_\lambda)\), so \(F_{jk}\) reduces to the covariance matrix of the restrained observables. This derivation of the Fisher information from the KL divergence, and its subsequent use as a metric on the space of statistical models, is well established in information geometry~\cite{amari2000methods}. This use of the KL divergence is related, but not equivalent, to its role in the maximum-likelihood interpretation of MaxEnt. In the latter, one minimizes \(D_{\mathrm{KL}}[P^{\mathrm{exp}}\mid P_{\vec\lambda}]\) with respect to \(\vec\lambda\), while holding the experimental distribution \(P^{\mathrm{exp}}\) fixed~\cite{goodfellow2016deep,cesari2018using}. Here, instead, the KL divergence is evaluated between two members of the model family at neighboring parameter values, \(P_{\vec\lambda+d\vec\lambda}\) and \(P_{\vec\lambda}\), and its quadratic term in \(d\vec\lambda\) defines the Fisher information metric.

This construction is local: the Fisher metric follows from the second-order expansion of the
Kullback-Leibler divergence between infinitesimally close distributions. We use it here not as an assumption that all relevant refined ensembles remain close to the reference ensemble, but as a local volume element on the restricted manifold \(\{P_{\vec\lambda}\}\). This use of relative entropy, or equivalently of \(D_{\mathrm{KL}}\) up to sign convention, is independent of its role in Eq.~\ref{eqn:loss_ER}, where \(D_{\mathrm{KL}}\) enters as a regularization term penalizing deviations from the reference ensemble. Here, instead, the same local geometry is used only to define how posterior mass should be counted under changes of parametrization.

Notice that the derivation in Eq.~\ref{eqn:ds2_jeff} is quite general and can also be applied to force-field fitting, in which case the Fisher matrix is given by the covariance matrix of the generalized forces $f_j(\vec\phi)=\frac{\partial V(\vec\phi)}{\partial \phi_j}$, where $\phi$ indicates the set of force-field parameters free to vary. However, a bijective (one-to-one) correspondence between the set of parameters $\phi$ and the ensembles $P_\phi$ must still hold, which is ensured as long as the Fisher information matrix $F$ is positive definite.
This requires the coefficients $\phi$ to describe linearly independent corrections to the force field.

Assigning equal posterior volume to equal infinitesimal distances \(ds\) gives the local volume element around \(\vec\lambda\):
\begin{equation}
dn(\vec\lambda) = \sqrt{\det F(\vec\lambda)} \, d\vec\lambda
\label{eqn:dn}
\end{equation}
where the density $\sqrt{\det F(\vec\lambda)}$ is precisely the Jeffreys ensemble-counting measure.
The Jeffreys factor does modify the coordinate density used for sampling in $\lambda$-space. However, throughout this work, we define the MAP ensemble as the minimizer of the original loss function $\mathcal{L}[P]$; the ensemble-counting measure is introduced only for computing posterior averages and uncertainties.
Since our goal is to compute posterior averages, the relevant object is the posterior measure itself rather than the location of the maximum of a particular posterior density.
This measure should be invariant under reparametrization of the same ensemble manifold and normalizable, so that Bayesian averages and uncertainties are well defined.

\begin{equation}
    d\mathcal P(P_{\vec\lambda})
\propto
e^{-\mathcal L[P_{\vec\lambda}]}
\sqrt{\det F(\vec\lambda)}\,d\vec\lambda .
\end{equation}

Equivalently, when sampling in the coordinates \(\vec\lambda\), this corresponds to the effective loss
\begin{equation}
\mathcal L'[P_\lambda] = \mathcal L[P_\lambda] - \log\sqrt{\det F(\lambda)}.
\end{equation}

\begin{figure*}
\centering
\subfloat[]{\includegraphics[width=0.5\textwidth]{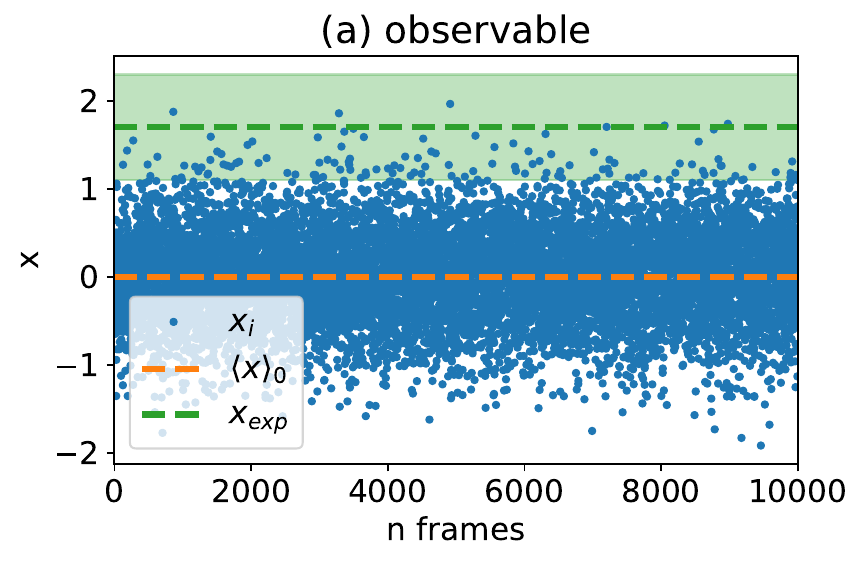}}
\subfloat[]{\includegraphics[width=0.5\textwidth]{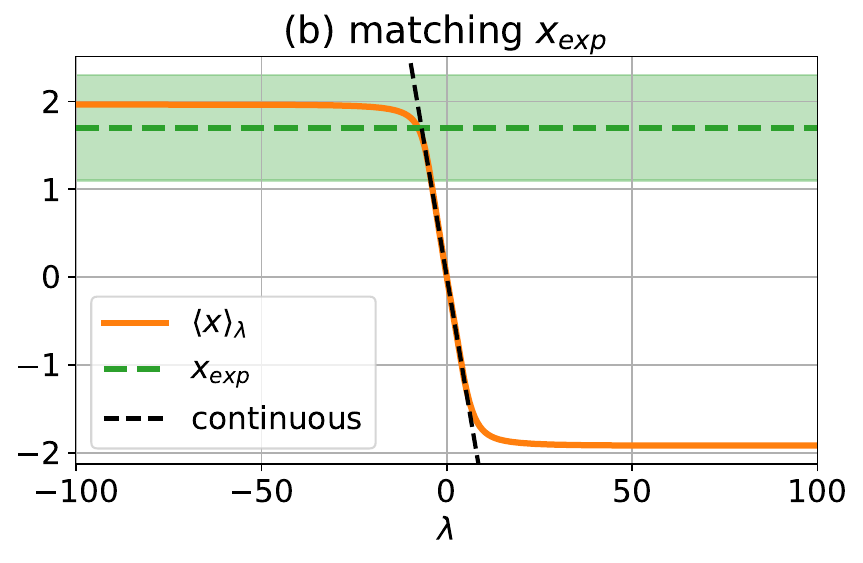}}
\hfill
\subfloat[]{\includegraphics[width=0.5\textwidth]{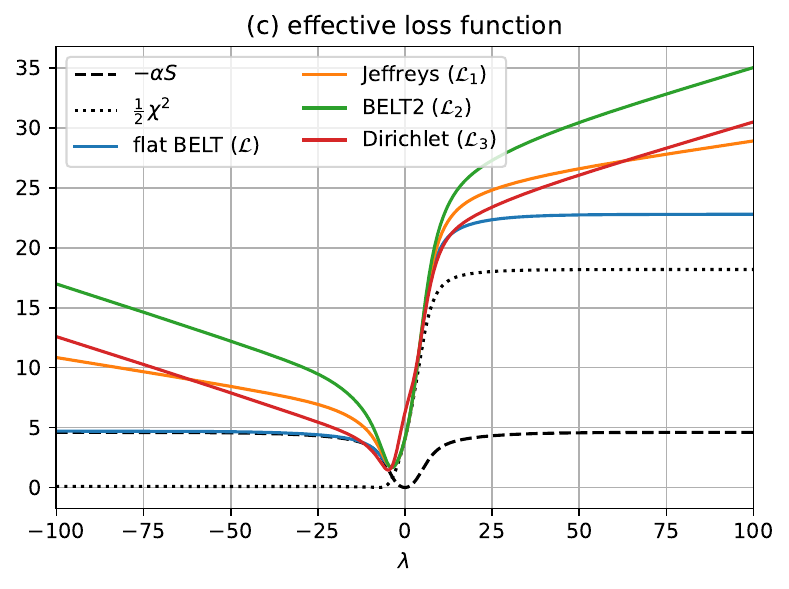}}
\subfloat[]{\includegraphics[width=0.5\textwidth]{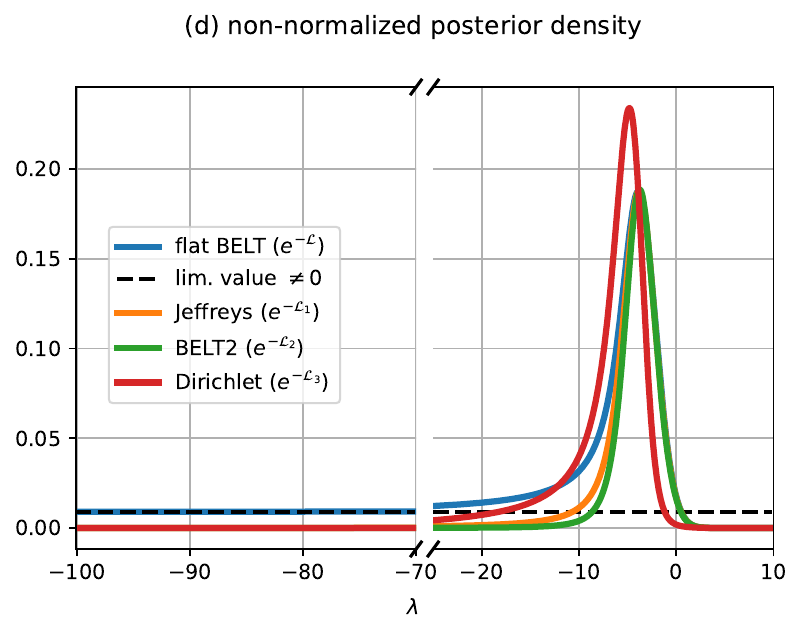}}
\caption{(a)
Values of $x$ sampled from the Gaussian reference distribution $x\sim N(0;\sigma^2)$, with $\sigma=0.5$,
 together with the experimental measurement $x_{exp}=1.7\pm0.6$.
(b) Reweighted average $\langle x \rangle_{\lambda}$ as a function of the parameter $\lambda$.
The experimental value is matched for $\lambda\simeq -7.7$.
For large $|\lambda|$, the average approaches the largest or smallest sampled value and becomes independent of $\lambda$.
(c) Flat-BELT loss function $\mathcal L = \frac{1}{2}\chi^2 - \alpha S_{rel}$, for $\alpha=0.5$, and
the corresponding effective losses obtained by including the Jeffreys ($\mathcal{L}_1$),
BELT2 ($\mathcal{L}_2$),
and Dirichlet ($\mathcal{L}_3$) ensemble-counting measures.
(d) Corresponding unnormalized posterior densities. The flat BELT measure produces a finite asymptotic plateau (dashed line), whereas the alternative measures suppress the plateau region and yield normalizable posteriors.
}
\label{fig:example1}
\end{figure*}

\subsection{BELT2 measure}

An alternative ensemble-counting measure follows from the formulation proposed in a preprint by Lane et al.\cite{lane2014efficient}, where the posterior is first defined in terms of the restrained observable averages and then transformed to the \(\vec{\lambda}\) coordinates. The resulting Jacobian is given by the determinant of the covariance matrix of the restrained observables. We refer to this covariance-determinant measure as ``BELT2''. To our knowledge, this terminology was not used in the original work and is adopted here only for convenience. This construction is equivalent to sampling uniformly in the space of restrained observable averages \(\langle \vec g \rangle_{\vec\lambda}\), provided that the mapping between \(\langle \vec g \rangle_{\vec\lambda}\) and \(\vec\lambda\) is one-to-one, as is the case for independent observables.
Then, the infinitesimal distance squared between two ensembles with average values $\langle \vec g \rangle_\lambda$ and $\langle \vec g \rangle_\lambda + d\langle \vec g\rangle_\lambda$ is

\begin{equation}
\begin{split}
ds^2 & = \sum_i d\langle g_i \rangle_\lambda^2
= \sum_j \Bigl(\sum_k \frac{\partial\langle g_j\rangle_\lambda}{\partial\lambda_k} d\lambda_k \Bigr)^2
\\ & = \sum_j \Bigl(-\sum_k C_{jk}(\vec\lambda) d\lambda_k \Bigr)^2 \\
& = \sum_j \sum_{k,k'} C_{jk}(\vec\lambda) C_{jk'}(\vec\lambda) d\lambda_k d\lambda_k'
\label{eqn:ds2_av}
\end{split}
\end{equation}
where $C(\vec\lambda)$ 
is the covariance matrix of the observables, as defined in the previous Subsection. Then, the local number of ensembles in a volume $d\lambda$ centered around $\vec\lambda$ is given by the volume transformation as 
\begin{equation}
dn(\lambda) = \sqrt{\det C^T(\vec\lambda) C(\vec\lambda)} \, d\vec\lambda = \det C(\vec\lambda) \, d\vec\lambda.
\end{equation}

However, this measure is defined through the chosen observable averages $\langle \vec g\rangle$ rather than through a metric on the ensembles themselves. While the BELT2 choice resolves the normalization issue of the flat-$\lambda$ posterior, it assigns local volume according to $\det C(\vec\lambda)$, rather than according to $\sqrt{\det C(\vec\lambda)}$ as in the Jeffreys construction. The two measures can therefore assign different posterior mass when the covariance varies across the ensemble manifold.

\subsection{Dirichlet measure}

In Ref.~\cite{hummer2015bayesian},
the posterior was defined as a function of the weights
of each frame $\mathcal{P}[w_1,\cdots,w_N]$, and sampled uniformly on the unit simplex defined by positive normalized weights. This corresponds to measuring distances between ensembles using the Euclidean metric in weight space. Since the uniform measure on the simplex is $\mathrm{Dir}(\alpha=1)$, we will refer to this choice as the ``Dirichlet’’ measure.
We here adjust this definition to the case of
ensembles parametrized by the $\vec\lambda$ coefficients as in the previous cases (Eqs. \ref{eqn:ds2_jeff}, \ref{eqn:ds2_av}). The infinitesimal distance squared becomes:
\begin{equation}
\begin{split}
ds^2  & = \sum_x (d P(x))^2 = \sum_x \Bigl(\sum_j \frac{\partial P(x)}{\partial \lambda_j} d\lambda_j\Bigr)^2  \\ &= \sum_{j,k} M_{jk}(\lambda) \, d\lambda_j d\lambda_k
\end{split}
\end{equation}
where
\begin{equation}
\begin{split}
M_{jk}(\lambda) & = \sum_x \frac{\partial P_\lambda(x)}{\partial \lambda_j} \frac{\partial P_\lambda(x)}{\partial\lambda_k} \\
& = \sum_x \Bigl [g_j(x) g_k(x) + \langle g_j \rangle_\lambda \langle g_k \rangle_\lambda  \\ & - (\langle g_j \rangle_\lambda \, g_k(x) + %
\langle g_k \rangle_\lambda \, g_j(x))
\Bigr] P_\lambda(x)^2
\label{eqn:metric_dirichlet}
\end{split}
\end{equation}
The corresponding local volume element on the restricted \(\vec\lambda\) manifold is therefore
\[
dn_{\mathrm{Dir}}(\vec\lambda)
=
\sqrt{\det M(\vec\lambda)}\,d\vec\lambda .
\]
The main difference with the Jeffreys measure is that we are now focusing on absolute variations $\delta P(x)^2$ rather than relative ones $\delta P(x)^2/P(x)$. As a consequence, changes in low-probability frames contribute less to the metric than one would reasonably expect.
For instance, a relative 10\% increment in the weight of a simulation frame with population
1\% contributes less than the same relative increment in the weight of another simulation frame
with population 50\%.
The induced ensemble density is therefore not determined solely by the distinguishability between nearby ensembles, but also by how probability mass is distributed among frames. In particular, the density becomes small both when the ensemble is concentrated on a few frames and when the weights are nearly uniform, whereas it is larger in regions where probability is redistributed more unevenly among competing frames.
The Euclidean metric in weight space corresponds to the uniform measure on the simplex, i.e. $\mathrm{Dir}(\alpha=1)$, whereas the Fisher metric corresponds to $\mathrm{Dir}(\alpha=1/2)$. We do not pursue this analogy further because the present work focuses on the restricted manifold $P_\lambda$.

\section{Results}

\begin{figure*}
\centering
\subfloat[]{\includegraphics[width=0.5\textwidth]{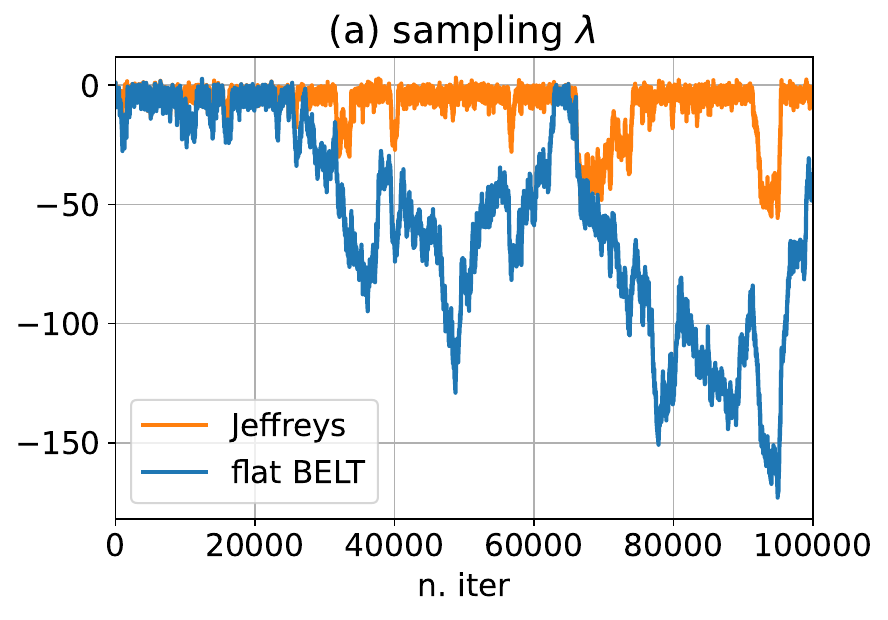}}
\subfloat[]{\includegraphics[width=0.5\textwidth]{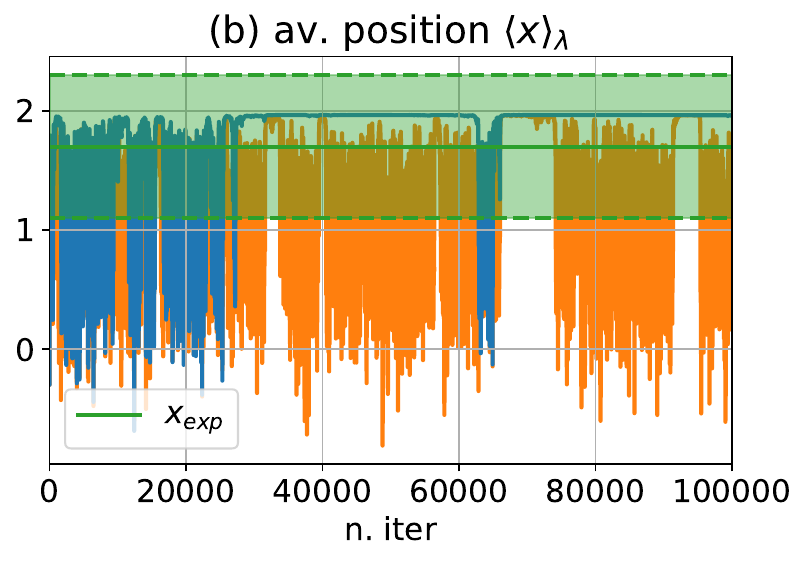}}
\hfill
\subfloat[]{\includegraphics[width=0.5\textwidth]{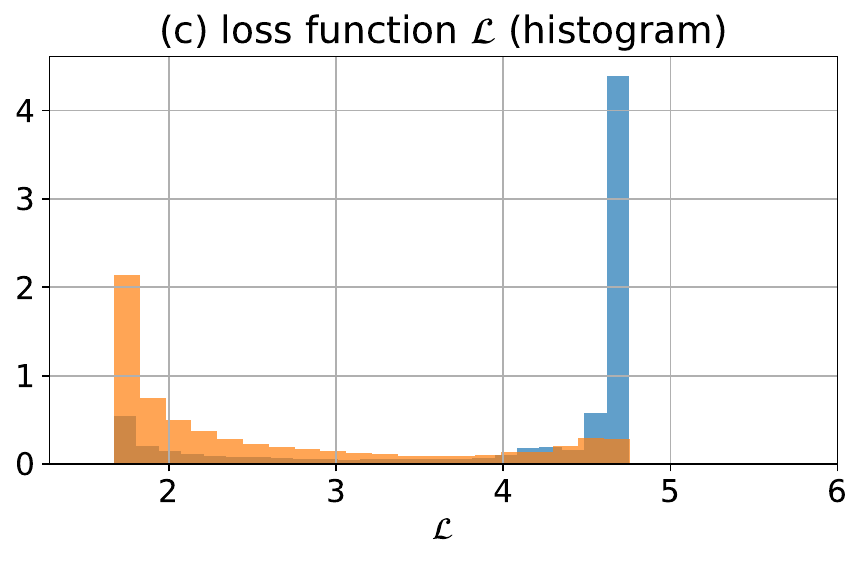}}
\subfloat[]{\includegraphics[width=0.5\textwidth]{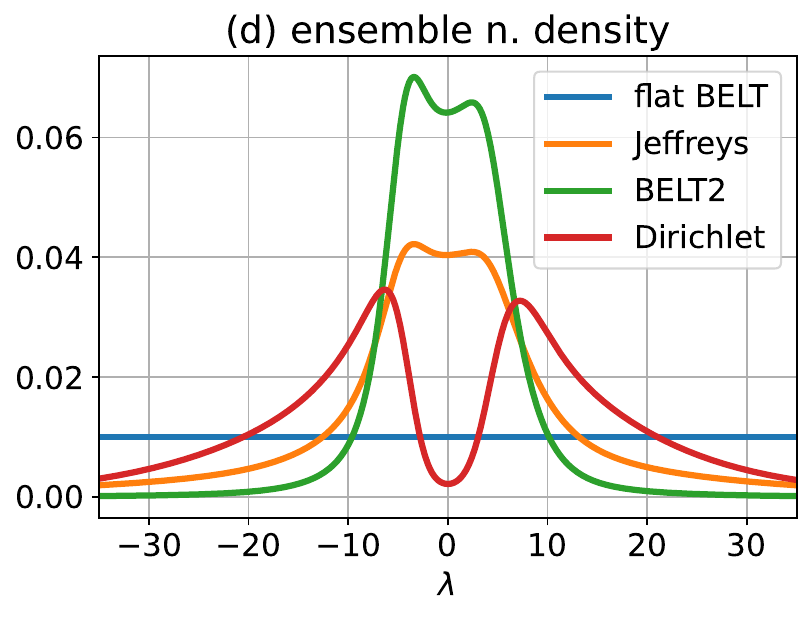}}
\caption{(a,b) Metropolis sampling of the posterior distribution shown in Fig.~1, performed with the flat BELT and Jeffreys ensemble-counting measures. With flat BELT, the sampler occasionally becomes trapped in the plateau regions, visible as horizontal segments in panel (b). These excursions bias the posterior average of $\langle x\rangle_\lambda$. (c) Histogram of the original loss function $\mathcal L$ (excluding ensemble-counting contributions) obtained from the two samplings. (d) Ensemble-counting densities induced by the flat BELT, Jeffreys, BELT2, and Dirichlet measures, computed using the same finite Gaussian sample as in Fig.~1. The $\lambda$ dependence and asymmetry of the nonuniform densities arise from this finite sampled representation and would disappear if the corresponding analytical continuum Gaussian distribution were considered.}
\label{fig:example1_sampling}
\end{figure*}

\subsection{Finite-sampling effects in a Gaussian model}
\label{sec:quadratic}

We first consider a model system with a single degree of freedom subject to a quadratic potential
$V(x)=\frac{x^2}{2\sigma^2}$ (Section~\ref{sec:escaping}). Assuming the thermal energy $k_BT=1$, this corresponds to drawing $x$ 
from a normal distribution $x\sim\mathcal{N}(0,\sigma^2)$.
We then introduce an experimental measurement $\langle x\rangle = x_{exp}\pm\sigma_{exp}$
(see Fig.~\ref{fig:example1}a).
In the continuum limit, maximum-entropy reweighting
of a Gaussian reference distribution gives another 
Gaussian 
with the same variance and mean shifted by an amount proportional to the parameter $\lambda$.
For a finite trajectory, however,
the reweighted average is bounded by the largest and smallest sampled values of $x$
(see Fig.~\ref{fig:example1}b).
Consequently, for sufficiently large $|\lambda|$,
the reweighted ensemble becomes dominated by a single sampled frame.
In this regime, the ensemble becomes essentially independent of $\lambda$.
The loss function \(\mathcal L=\chi^2/2-\alpha S\) therefore approaches finite limiting values as \(\lambda\to\pm\infty\) (Fig.~\ref{fig:example1}c). This does not affect the MAP estimate, but it makes the flat-\(\lambda\) posterior used in BELT non-normalizable, because \(\mathcal P(\lambda)\propto e^{-\mathcal L(\lambda)}\) does not decay to zero at large \(|\lambda|\) (Fig.~\ref{fig:example1}d).

We then compared different choices of ensemble-counting measure on the same restricted family \(P_\lambda\). Including the Jeffreys measure introduces a density factor proportional to the local statistical distinguishability of neighboring ensembles, which vanishes in the plateau regions.
As a result, the posterior density goes to zero for large \(|\lambda|\), restoring normalizability.
A similar, although not identical, behavior is obtained by defining volume through the average value of the observable (BELT2) or through the Euclidean metric in weight space restricted to the \(P_\lambda\) manifold (Dirichlet measure). The resulting effective losses and posterior densities are shown in Figs.~\ref{fig:example1}c--d.

The practical consequence is visible in direct Monte Carlo sampling (Figs.~\ref{fig:example1_sampling}a--c). With the flat BELT measure, the sampler can spend long intervals in the \(\lambda<0\) plateau, where the ensemble is dominated by the frame with the largest sampled value of \(x\).
These intervals contribute a disproportionate number of posterior samples with a relatively large loss value,
here \(\mathcal L\simeq 4.7\), and alter the Bayesian posterior average of \(\langle x\rangle_\lambda\). By contrast, when the Jeffreys measure is included, the sampler remains localized in the region where the posterior mass is concentrated and avoids repeated sampling of the collapsed plateau ensemble.
Additional trajectories obtained with different random seeds are shown in Supplementary Fig.~\ref{fig:ex1_different_seeds}.

The induced ensemble densities for the four choices are shown in Fig.~\ref{fig:example1_sampling}d. These results were obtained from the finite Gaussian sample used in Fig.~\ref{fig:example1}a. Importantly, this finite sampled representation is also the origin of the asymmetry in panel (d): although the underlying Gaussian distribution is symmetric, the positive and negative tails of a particular finite random sample are not exact mirror images, so tilting in the two directions produces different covariance and density profiles. If the corresponding analytical continuum distribution with the linear observable $g(x)=x$ were considered instead, exponential tilting would change the mean but leave the variance independent of $\lambda$. The flat BELT, Jeffreys, and BELT2 densities would therefore be constant. The density associated with the continuum Euclidean metric underlying the Dirichlet construction would also be constant because this metric is invariant under translations of the Gaussian density. Consequently, all four prescriptions would differ at most by a constant factor and lead to the same normalized posterior in that continuum case.
Their dependence on the number of sampled frames is reported in Supplementary Fig.~\ref{fig:several_strides}, illustrating the increasing localization of the Jeffreys and BELT2 measures and the emergence of the central minimum in the Dirichlet measure.
The Jeffreys and BELT2 measures are closely related in this one-dimensional example, because both are determined by the variance of the restrained observable.
The Dirichlet measure instead exhibits a pronounced local minimum near $\lambda\simeq0$. This behavior is a finite-sampling effect arising from the discrete representation of the ensemble. Near $\lambda=0$, the weights are nearly uniform and therefore change only weakly under small variations of $\lambda$. As $|\lambda|$ increases, probability mass is redistributed among competing frames, leading to larger changes in the weights and a larger induced density. For sufficiently large $|\lambda|$, the ensemble becomes dominated by a small number of frames, and eventually by a single frame in the plateau region, so that the weights again become nearly insensitive to $\lambda$. Consequently, the induced density displays a local minimum near the reference ensemble and vanishes in the collapsed-ensemble plateaus.

\begin{figure}
\includegraphics[width=1\linewidth]{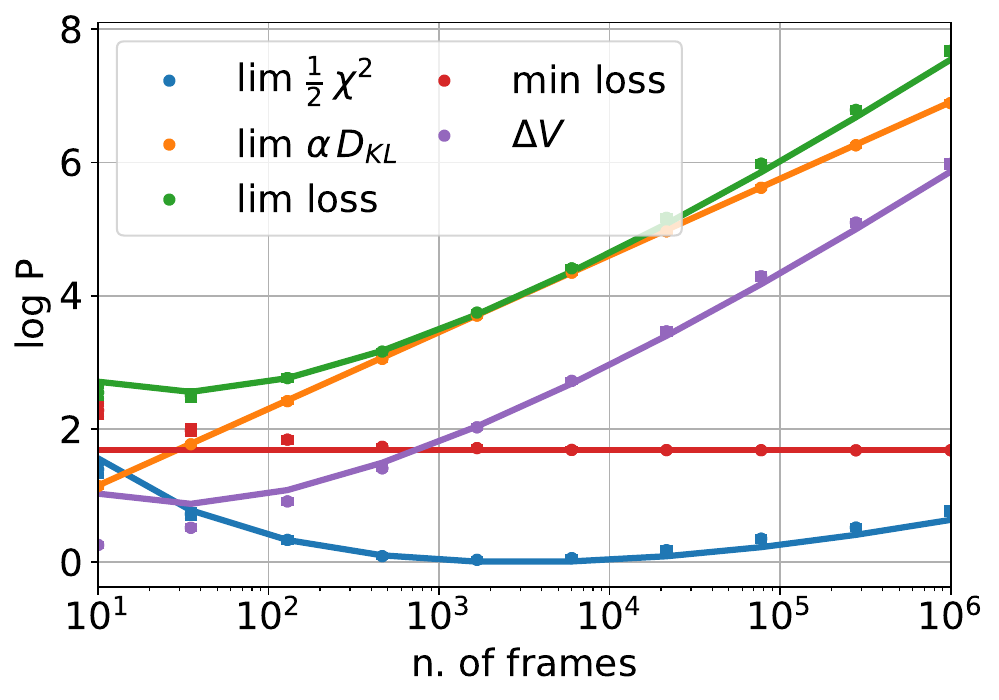}
\caption{
 Scaling of the quantities contributing to the loss difference $\Delta V$ between the minimum of the loss function and the lowest plateau, as a function of the number of frames. Symbols show averages over 50 random realizations of $\{g_i\}$, using the same values of $\alpha$, $g_{\rm exp}$, $\sigma_{\rm exp}$, and reference distribution as in Fig.~\ref{fig:example1}. Solid lines show the analytical scaling laws derived in Eqs.~\ref{eqn:lambda_star} and \ref{eqn:scaling_losslim}. The logarithmic growth of $\Delta V$ originates primarily from the relative-entropy contribution associated with the collapsed plateau ensemble. Deviations at small numbers of frames reflect the breakdown of both the continuum approximation for the loss minimum and the leading large-$N$ estimate of the extreme sampled frame defining the plateau.}
\label{fig:scaling}
\end{figure}

The severity of the flat-BELT non-normalizability depends on the loss difference between the region around the minimum of the loss function and the lowest plateau. For a finite Monte Carlo run, this difference controls how easily the sampler reaches the plateau region. In the quadratic model, the plateau corresponds to an ensemble dominated by a single sampled frame, so its relative entropy with respect to the uniform discrete reference ensemble depends explicitly on the number of frames. Figure~\ref{fig:scaling} reports the scaling of the different contributions to this loss difference as the number of frames is varied. The empirical estimates, obtained by averaging over multiple Gaussian samples, are in excellent agreement with the analytical scaling derived in Section~\ref{sec:escaping}. In particular, the loss difference grows logarithmically with the number of frames,
making plateau excursions progressively less likely.
Conversely, the artifact can become significant when the trajectory is strongly decimated. The deviations from the analytical curves at the smallest numbers of frames are expected because both the continuum approximation for the loss minimum and the leading large-$N$ estimate of the extreme sampled frame defining the plateau become less accurate.

\subsection{Bayesian refinement of an RNA oligomer}
\label{sec:rna-oligomer}

As a realistic test case, we considered maximum-entropy refinement of the structural ensemble of an (r)AAAA RNA oligomer using 28 \(^{3}J\) scalar couplings associated with the backbone \(\beta,\gamma,\epsilon,\nu\) dihedral angles.
The reference MD trajectory was taken from Ref.~\cite{frohlking2023simultaneous}.
The trajectory was generated using the
AMBER BSC0 OL3 RNA force field \cite{cornell1995second,wang2000well,perez2007refinement,zgarbova2011refinement} with van der Waals modification of phosphate oxygens \cite{steinbrecher2012revised}; water was modeled with the OPC model 
\cite{izadi2014building}. Sampling was enhanced using parallel tempering
\cite{sugita1999replica}. Experimental data were taken from Ref.~\cite{condon2015stacking}.
The scalar couplings were evaluated using the corresponding, distinct Karplus coefficient sets for the backbone \(\gamma\), backbone \(\beta/\epsilon\), and sugar-coupling forward models, as in Ref.~\cite{frohlking2023simultaneous}.

\begin{figure*}
\centering
\subfloat[]{\includegraphics[width=0.96\textwidth]{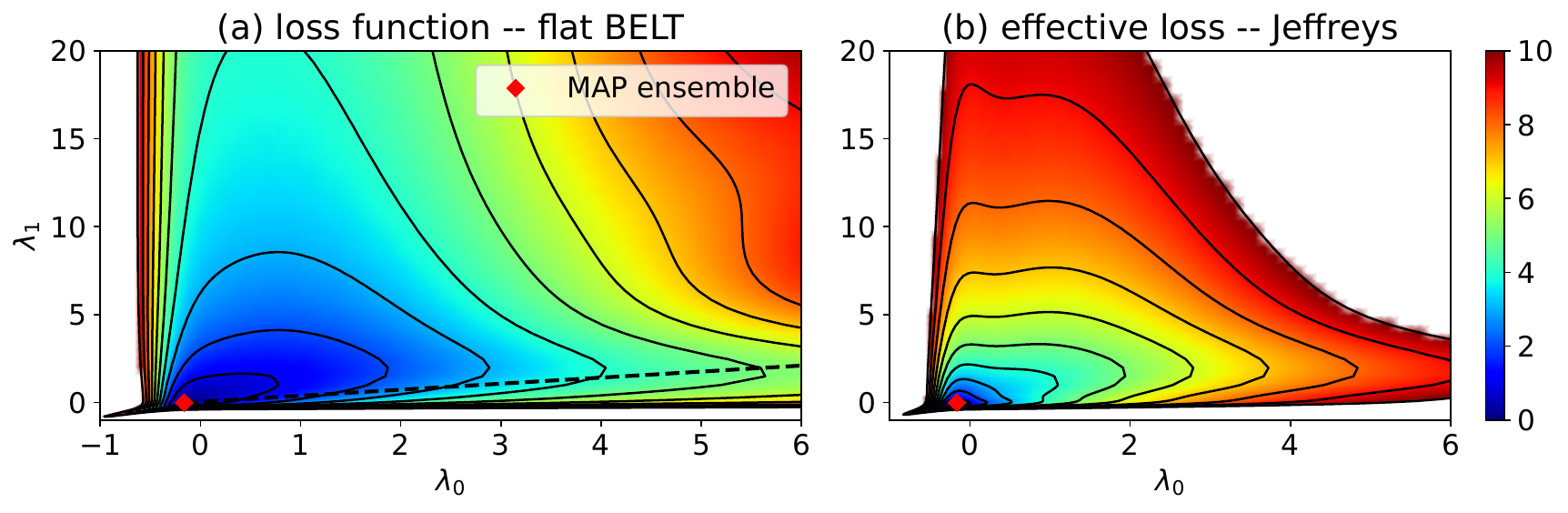}}
\hfill
\subfloat[]{\includegraphics[width=0.96\textwidth]{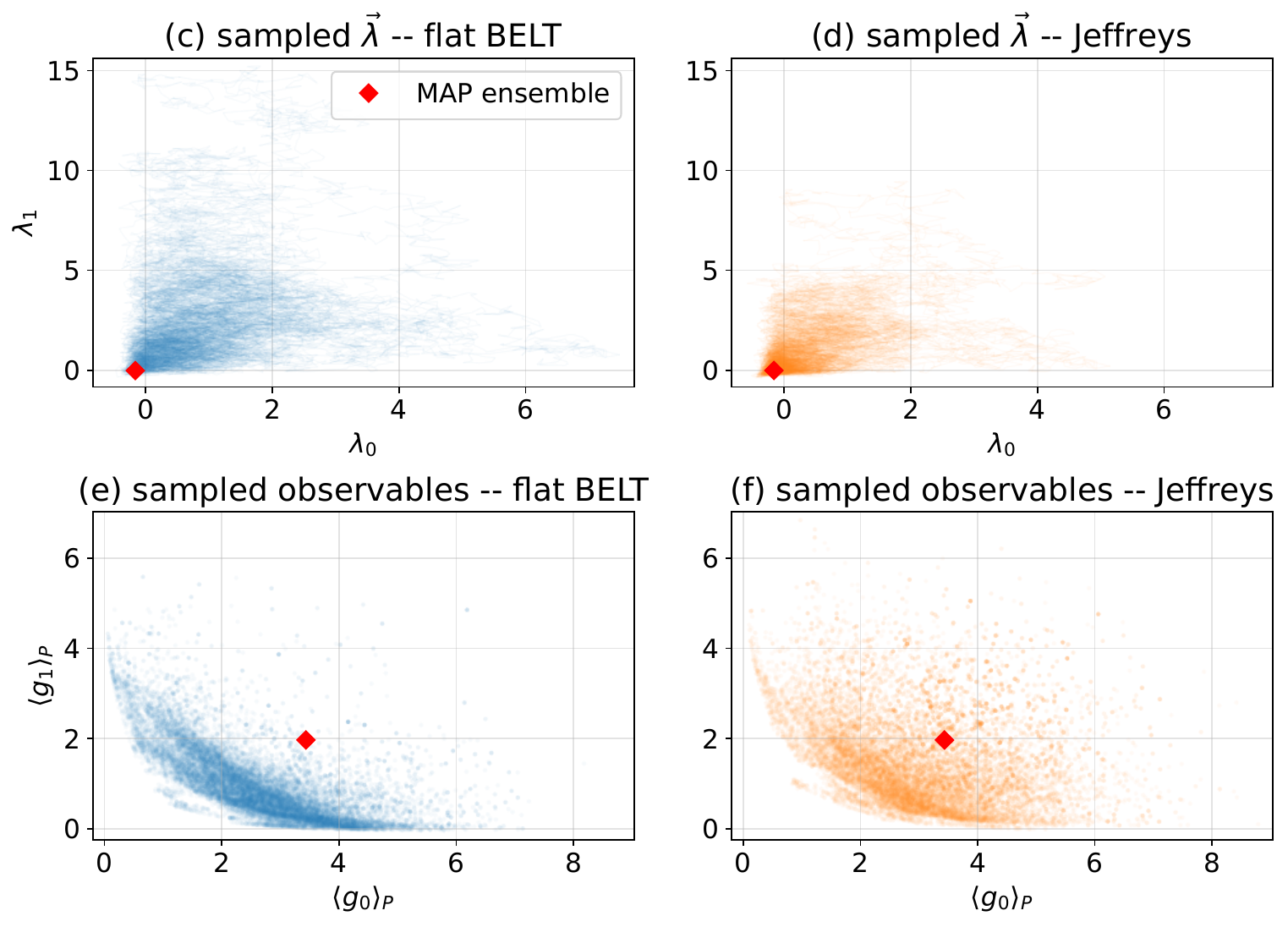}}
\caption{
Maximum-entropy refinement of the RNA oligomer using two $^{3}J$ scalar couplings ($\alpha=1$; stride =10, corresponding to 101,700 MD frames). (a) Loss function $\mathcal L(\vec\lambda)$ obtained with the flat BELT measure. The dashed line indicates the approximate direction of the lowest plateau. (b) Effective loss obtained by including the Jeffreys ensemble-counting measure. The position of the minimum of the original loss function (red diamond) is unchanged because the ensemble-counting measure does not affect the MAP estimate. (c,d) Representative Metropolis trajectories in $\vec\lambda$ space for the flat BELT and Jeffreys measures, respectively. Transparent trajectory segments are used so that repeatedly visited regions appear darker. The flat BELT posterior allows extensive exploration of the plateau region, whereas the Jeffreys measure concentrates sampling near the region containing most posterior probability mass. (e,f) Corresponding posterior samples of the restrained observables, also plotted with transparency to indicate their sampled densities.}
\label{fig:example2}
\end{figure*}

We first restrained only two experimental data points, corresponding to two \(^{3}J\) scalar couplings related to \(\gamma\) dihedral angles, so that the posterior could be visualized in two dimensions. Figures~\ref{fig:example2}a--b compare the original flat-BELT posterior with the posterior including the Jeffreys measure. With the flat measure, the posterior exhibits a plateau direction, approximately indicated by the dashed line in Fig.~\ref{fig:example2}a, whose lowest asymptotic value lies \(\Delta V\simeq9.8\) above the minimum. As a result, the Monte Carlo trajectory explores a broad region with large \(\lambda_0,\lambda_1\) values (Fig.~\ref{fig:example2}c). Including the Jeffreys measure suppresses this plateau and confines the sampling to the region containing most of the posterior probability mass (Fig.~\ref{fig:example2}d), leading to a different posterior distribution of refined observable averages (Figs.~\ref{fig:example2}e--f).
The difference between the two effective losses, $-\frac{1}{2}\log\det C(\vec\lambda)$ up to an additive constant, is shown directly in Supplementary Fig.~\ref{fig:example2_correction}. The correction varies by approximately six loss units over the displayed domain and increases toward the collapsed-ensemble plateau, making explicit how the Jeffreys measure suppresses that region.
The corresponding time series and marginal distributions of the sampled observable averages are reported in Supplementary Fig.~\ref{fig:example2_si}.
Additional two-dimensional posterior visualizations for a larger regularization hyperparameter are shown in Supplementary Fig.~\ref{fig:example2_alpha10}.

\begin{figure*}
\centering
\subfloat[]{\includegraphics[width=0.5\textwidth]{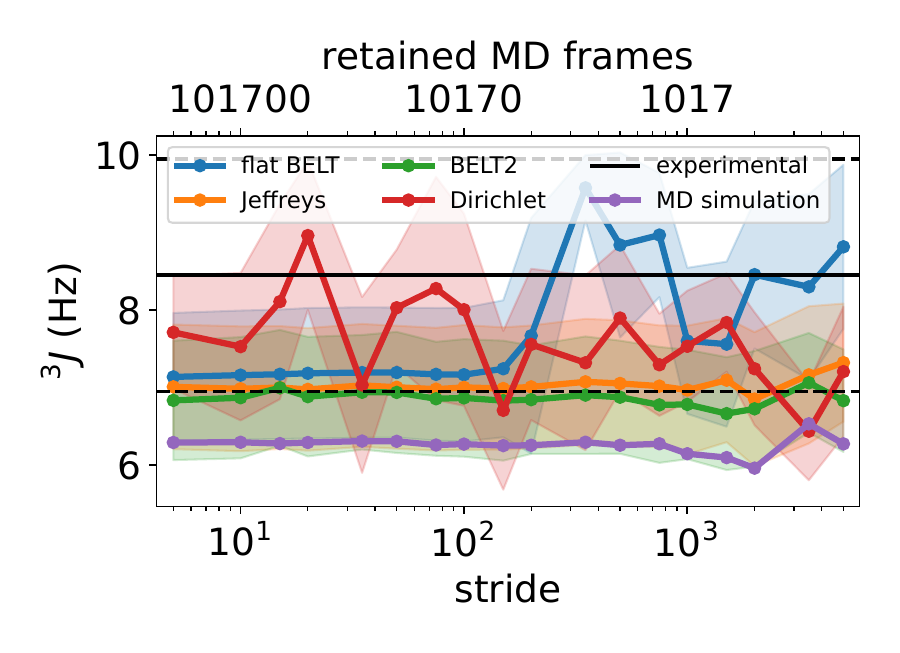}}
\subfloat[]{\includegraphics[width=0.5\textwidth]{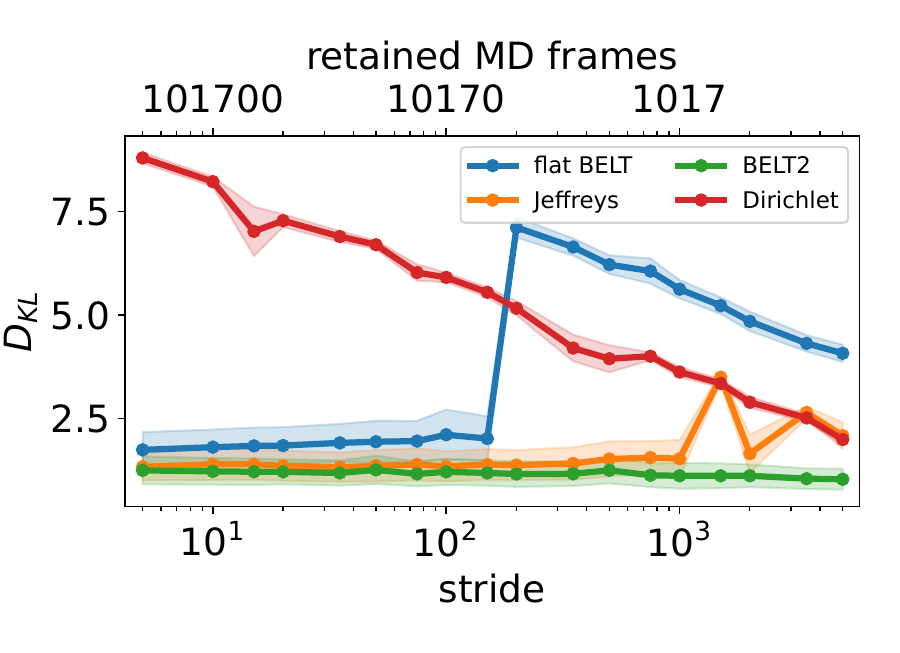}}
\caption{Posterior averages of observable 8 (left) and of the Kullback–Leibler divergence $D_{\mathrm{KL}}$ (right) obtained using the four ensemble-counting measures for different trajectory strides. Each curve was computed from the last 250,000 steps of a 500,000-step MCMC trajectory. Shaded regions show the posterior mean plus or minus one standard deviation and do not represent Monte Carlo standard errors. The upper axis reports the corresponding number of retained MD frames. BELT2 and, to a lesser extent, Jeffreys give the most stable estimates across strides. Flat BELT exhibits a marked dependence on the discretization of the reference ensemble, whereas for Dirichlet this dependence is clearer in $D_{\mathrm{KL}}$ than in observable 8; residual block-to-block variations indicate that some flat-BELT and Dirichlet calculations are not fully converged on this sampling timescale.}
\label{fig:example2b}
\end{figure*}

For the same system, we then restrained all 28 available \(^{3}J\) scalar couplings simultaneously and repeated posterior sampling for several strides of the MD trajectory, using the four ensemble-counting measures described above. The hyperparameter was set to \(\alpha=10\). Each MCMC trajectory comprised 500,000 steps; we discarded the first 250,000 steps and computed posterior means and standard deviations from the second half. These standard deviations quantify uncertainty across the posterior distribution of refined ensembles. They should not be confused with the Monte Carlo standard errors of the estimated posterior means, which concern only the numerical precision of the MCMC calculation and are not reported here. Comparisons between two consecutive 125,000-step blocks within the retained half showed stable estimates for BELT2 and for most Jeffreys calculations, whereas larger block-to-block variations remained for flat BELT and Dirichlet. Thus, the latter curves should be interpreted as finite-time sampling results rather than fully converged posterior estimates; for flat BELT, convergence is precluded in principle by the non-normalizable target.
The resulting dependence on trajectory stride provides a direct measure of the sensitivity of each ensemble-counting prescription to the finite representation of the reference ensemble (Fig.~\ref{fig:example2b} and Supplementary Material).

Figure~\ref{fig:example2b}a shows a representative observable, number~8, corresponding to the \(\beta\) dihedral angle \(C3^\prime(0)-O3^\prime(0)-P(1)-O5^\prime(1)\) between residues 0 and 1.
We focus on this observable because the unrefined MD average lies outside the experimental uncertainty, making the effect of ensemble refinement particularly evident.
Posterior averages obtained with BELT2 remain nearly constant over several orders of magnitude in trajectory stride. Jeffreys gives similarly stable averages for observable 8 and is substantially more stable than flat BELT and Dirichlet overall, although its posterior average of $D_{\mathrm{KL}}$ shows isolated deviations in the strongly decimated regime. With the flat BELT measure, the posterior average is close to the Jeffreys/BELT2 values for small strides, but changes abruptly between strides of approximately $10^2$ and a few times $10^2$, consistently with plateau regions becoming increasingly accessible during sampling. The same transition is visible in the posterior average of $D_{\mathrm{KL}}$ in Fig.~\ref{fig:example2b}b. For observable 8, the flat-BELT estimate moves from below the experimental central value to values above it at some strides, but this should not be interpreted as a systematic improvement or deterioration of agreement with experiment. Since all 28 observables are restrained simultaneously, the plateau explored by the sampler is determined by the joint loss rather than by the residual of observable 8 alone. The concurrent discontinuities in the observable and $D_{\mathrm{KL}}$ reflect sampling of the pathological plateau region of the finite reference ensemble.
For observable 8, the Dirichlet posterior mean fluctuates with stride without a simple monotonic trend, and these variations are generally modest compared with its posterior standard deviation. Figure~\ref{fig:example2b}a reports only this observable; results for all restrained observables, which show varying degrees of stride dependence, are provided in Supplementary Figs.~\ref{fig:other_obs1} and~\ref{fig:other_obs2}.

The posterior average of \(D_{\mathrm{KL}}\) provides a more sensitive view of the dependence on trajectory stride (Fig.~\ref{fig:example2b}b). In particular, with the Dirichlet measure it decreases overall as the stride increases, whereas the corresponding changes in the posterior mean of observable 8 show no analogous monotonic trend and are modest compared with its posterior standard deviation. This is not contradictory: a single restrained average probes only one projection of the refined ensemble, and similar values of that average can be produced by ensembles with substantially different redistributions of statistical weight and hence different \(D_{\mathrm{KL}}\). The Dirichlet trend in \(D_{\mathrm{KL}}\) reflects the reduced density near the reference ensemble when the number of frames is large, as also visible in Supplementary Fig.~\ref{fig:several_strides}. BELT2, and to a lesser extent Jeffreys, are considerably more stable under trajectory decimation. With the flat BELT measure, the posterior average of \(D_{\mathrm{KL}}\) remains close to the Jeffreys/BELT2 results at lower strides but changes abruptly at larger strides, consistently with the finite-sample plateau becoming accessible during sampling.

Overall, the realistic RNA example confirms the conclusions of the Gaussian model. Posterior averages depend significantly on the choice of ensemble-counting measure, with BELT2 and Jeffreys substantially more stable under trajectory decimation than flat BELT and Dirichlet. These results indicate that Bayesian estimates obtained with the flat BELT and Dirichlet measures can be strongly affected by the discretization used to represent the reference ensemble.

\section{Discussion}

In summary, this study investigates the consequences of the choice of ensemble-counting measure when Bayesian ensemble refinement is extended beyond the commonly used maximum-a-posteriori (MAP) estimate. Most applications of ensemble refinement
focus on the MAP ensemble obtained by minimizing a loss function such as Eq.~\ref{eqn:loss_ER}.
Full posterior sampling, however, can provide uncertainty estimates for refined observables and can reveal whether alternative predictions derived from different refined ensembles remain statistically plausible. Thus, posterior averages are not determined by the loss function alone: they also require specifying how volume is measured in the space of ensembles being sampled.

Several sampling strategies have been proposed in the literature~\cite{beauchamp2014bayesian,lane2014efficient,hummer2015bayesian}. These strategies differ not only in their parametrization, but also in the measure used to count ensembles. Bayesian sampling of weights is one possible approach~\cite{hummer2015bayesian,tang2023ensemble}, but posterior estimates then depend on the number of discrete states used to represent the system. For instance, saving a trajectory at higher frequency increases the dimension of the weight vector and can affect posterior averages and variances, even when the underlying simulation is unchanged. This issue is related to the difficult problem of estimating the number of independent configurations in a molecular simulation. In the present work, we instead restrict posterior sampling to the family of maximum-entropy ensembles \(P_{\vec\lambda}\), as proposed in BELT~\cite{beauchamp2014bayesian}. The Dirichlet measure tested here should not be confused with the weight-space sampling of Ref.~\cite{hummer2015bayesian}, where weights on individual snapshots are sampled without the \(\vec\lambda\) parametrization.

The ensemble-counting ambiguity is not a feature of every Bayesian formulation of ensemble refinement. For example, Bayesian Inference of Conformational Populations (BICePs) samples conformational states, or sets of replicas, together with nuisance parameters directly in configuration space and uses the MBAR estimator to compute a Bayesian-evidence score~\cite{raddi2025model,raddi2026automatic}. Because the force-field distribution supplies the base measure in configuration space, this formulation does not encounter the specific non-normalizability caused by a flat measure on an unbounded maximum-entropy $\vec\lambda$ manifold. The two approaches nevertheless quantify different objects: variability among configurations primarily represents conformational heterogeneity, whereas a posterior over $\vec\lambda$ explicitly represents uncertainty among alternative refined ensemble distributions. BICePs therefore provides an alternative Bayesian formulation rather than a different ensemble-counting measure on the same manifold. Extensions of this evidence-based framework have also been applied to forward-model and force-field parameterization~\cite{raddi2025model,raddi2026automatic}.

A restricted sampling strategy was introduced in BELT and BELT2~\cite{beauchamp2014bayesian,lane2014efficient}. This approach first selects, for each set of observable averages, the ensemble that maximizes the relative entropy with respect to the reference ensemble, and then samples this restricted family, parametrized by the coefficients \(\vec\lambda\). However, a posterior distribution on this family still requires a choice of measure. We have shown here that the simplest choice, namely a uniform measure in \(\vec\lambda\) as used in BELT~\cite{beauchamp2014bayesian}, leads to a non-normalizable posterior for finite samples. The reason is that, for large values of \(|\vec\lambda|\), the loss function reaches plateau regions where the refined ensemble is dominated by one or a few frames. With respect to the flat \(d\vec\lambda\) measure, the posterior density therefore remains finite over an unbounded region of parameter space, even though the corresponding ensembles change only minimally. Although this behavior is a direct consequence of representing the reference ensemble with a finite number of frames, we are not aware of it being previously reported.
These plateau regions should not be interpreted as physically meaningful refined ensembles. They correspond to extreme reweighting and would normally be regarded as non-gentle refinements \cite{kofinger2024encoding}. Their relevance here is methodological: a flat measure in \(\vec\lambda\) assigns nonzero posterior volume to extended regions that effectively represent the same collapsed ensemble. As a consequence, posterior averages can become sensitive to arbitrary choices such as the length of the Monte Carlo trajectory, imposed bounds on \(\vec\lambda\), or the number of frames used to discretize the reference ensemble. This also explains why diagnostics such as the relative entropy, effective sample size, or related measures of reweighting severity remain important when interpreting posterior samples.

The increase of the loss difference between the plateau and the minimum
with the number of sampled frames may explain why this issue has received limited attention so far. In long trajectories, the plateau can be separated from the region around the minimum of the loss function by a large loss difference, making it unlikely to be reached in practical Monte Carlo sampling. However, this separation depends on the number of analyzed frames and on the choice of \(\alpha\) and experimental uncertainty. As shown by our analysis on realistic molecular dynamics simulations, the problem can become visible when the trajectory is decimated or when the posterior is broad.

There is no universal threshold above which the flat BELT and Jeffreys measures can be considered equivalent. For any finite reference ensemble, the flat BELT posterior remains formally non-normalizable and a sufficiently long Monte Carlo trajectory may eventually reach the plateau region, preventing convergence of posterior estimates. The loss difference between the minimum and the plateau provides a qualitative indication of how readily this may occur, but the probability of observing a plateau excursion in a finite calculation also depends on the sampling protocol, including the proposal step size, the number of Monte Carlo steps, and the initial condition. Consequently, apparent agreement between flat BELT and Jeffreys estimates in a finite run does not establish their equivalence. If the flat BELT measure is used, the time series of the Lagrange multipliers, loss, and restrained averages should therefore be monitored explicitly for plateau excursions. A more robust solution is to employ a normalizable ensemble-counting measure, such as the Jeffreys measure proposed here.

The Jeffreys measure provides a natural way to define posterior volume on the restricted ensemble manifold. It is induced by the Fisher metric, is invariant under reparametrization, and suppresses the artificial posterior volume associated with collapsed plateau regions in finite samples. %

It is interesting to note that the Jeffreys prior was already considered among possible theoretical options in the supplementary information of the original BELT work \cite{beauchamp2014bayesian}, where it was not adopted because its maximum does not necessarily occur at the reference ensemble. This is a valid concern if the Jeffreys prior is intended to regularize the refinement toward \(\vec\lambda=0\). Our use is different: the regularization toward the reference ensemble is included in the loss function,
modulated by a suitably chosen hyperparameter, whereas the Jeffreys factor is used as an invariant volume element
and defines how posterior mass is counted when computing Bayesian averages.

BELT2, where sampling is performed uniformly over the average observable values \(\langle\vec g\rangle_P\), also gives a normalizable posterior for finite samples, since the observable averages are bounded. The Jeffreys and BELT2 densities are $\sqrt{\det C(\vec\lambda)}$ and $\det C(\vec\lambda)$, respectively. They therefore have the same computational overhead, namely the calculation of the covariance matrix and its determinant. They define the same normalized posterior when $\det C$ is a positive constant and give similar results when it varies little over the region containing most posterior mass. If the corresponding analytical continuum Gaussian distribution were considered instead of the finite sample used here, all four ensemble-counting densities would be constant in $\lambda$ and hence would give identical normalized posteriors. The differences visible in Fig.~\ref{fig:example1_sampling}d arise from the finite sampled representation of that Gaussian model. More generally, Jeffreys and BELT2 can diverge when $\det C(\vec\lambda)$ varies appreciably over the posterior uncertainty range in $\vec\lambda$. This is more likely when that range is broad, when the restrained observables have strongly non-Gaussian distributions (such as multimodal or bounded distributions), or when finite-sample effects make collapsed ensembles appreciably accessible; in such regions BELT2 suppresses small determinants more strongly than Jeffreys.

In the realistic example based on the reweighting of RNA oligomer trajectories, the choice of ensemble-counting measure had a visible practical effect. The flat BELT and Dirichlet measures produced posterior quantities that depended on the number of analyzed frames, which is undesirable because the saving frequency of a trajectory is often dictated by technical considerations such as storage requirements rather than by statistical independence. For the Dirichlet measure, this dependence is most evident in the Kullback--Leibler divergence. It is modest compared with the posterior uncertainty for observable 8 shown in Fig.~\ref{fig:example2b}a, while the complete set of restrained observables reported in Supplementary Figs.~\ref{fig:other_obs1} and~\ref{fig:other_obs2} shows varying degrees of stride dependence. More generally, similar values of an individual restrained average may result from different distributions of statistical weights and hence different Kullback--Leibler divergences. By contrast, Jeffreys and BELT2 gave posterior averages and Kullback--Leibler divergences that were more stable under trajectory decimation. This robustness supports the use of covariance-based measures, and in particular of the Jeffreys measure, for posterior sampling in ensemble refinement.

The hyperparameter \(\alpha\) should not be regarded as a purely numerical parameter. It controls the confidence assigned to the reference ensemble relative to the experimental data: small values allow stronger reweighting, whereas large values keep the refined ensemble closer to \(P_0\).
A cross-validation procedure assessing the capability of the procedure to generalize to
observables not used in training can be used to moderate overfitting
\cite{bottaro2018conformational}.
Other possible strategies include analysis of the \(\chi^2\)--\(D_{\mathrm{KL}}\) curve
\cite{rozycki2011saxs,kofinger2019efficient},
estimation of the statistical error of the simulation \cite{borthakur2025determining},
or the criterion that refinement should remain gentle~\cite{kofinger2024encoding}, which explicitly controls the statistical significance of the reweighting.
Beyond its role in controlling the strength of refinement, the hyperparameter $\alpha$ affects the practical impact of the non-normalizability of the posterior in the BELT method: for sufficiently large $\alpha$, the problematic regions of parameter space become effectively inaccessible, whereas for smaller values they can contribute significantly to posterior averages.

The same considerations discussed in this work are relevant for on-the-fly maximum-entropy restraining methods. Reweighting is reliable only when the reference simulation has sufficient overlap with the ensemble supported by the experimental data~\cite{shen2008statistical,rangan2018determination}.
If this is not the case, maximum-entropy restraints should be applied during the simulation, for example through time-dependent~\cite{white2014efficient,white2015designing,cesari2016combining} or replica-averaged schemes~\cite{lindorff2005simultaneous,cavalli2013molecular,roux2013statistical}. Error modeling can also be incorporated in these frameworks~\cite{cesari2016combining,bonomi2016metainference}. However, if such on-the-fly approaches are extended from MAP refinement to full Bayesian posterior sampling, the choice of posterior measure discussed here remains relevant. One should then distinguish two sources of variability: conformational sampling within each biased ensemble, and posterior sampling over alternative refined ensembles or biasing parameters.

As a perspective, as discussed in Refs.~\cite{kofinger2021empirical,gilardoni2024boosting}, there is a close analogy between ensemble refinement and force-field refinement.
While most existing approaches to force-field refinement focus on identifying optimal parameter values, Bayesian formulations based on posterior sampling of force-field parameters have also been proposed \cite{dutta2018bayesian,madin2022bayesian,sose2024evaluation}.
In that context, the Jeffreys measure is particularly attractive because it defines local volume through the statistical distinguishability of neighboring ensembles. Although not tested here, this approach has been implemented in the \texttt{MDRefine} Python package and is described in the Supplementary Material.

\section*{Supplementary Material}

The Supplementary Material contains implementation details of the methods described in this work, including their implementation in the \texttt{MDRefine} package. Additional Monte Carlo sampling results and ensemble-density analyses are provided for the Gaussian model. Supplementary posterior analyses are presented for the RNA oligomer refinement example, including observable distributions, dependence on the regularization parameter, and results for all restrained observables.

\begin{acknowledgments}
We thank J\"urgen K\"ofinger and Julija Zavadlav for valuable comments on an
early version of this work, presented as a chapter of Ivan Gilardoni's
doctoral thesis.
G.B. acknowledges support from the Italian Ministry of University and Research (MUR) through the Fondo Italiano per la Scienza (FIS 3), Grant No. FIS-2024-01745, RNAScale.

\end{acknowledgments}

\section*{Data Availability Statement}

The molecular simulation data analyzed in this work were previously published and are available from the sources cited in Refs.~\cite{frohlking2023simultaneous,gilardoni2025mdrefine}. All scripts and Jupyter notebooks required to reproduce the analyses, figures, and numerical results reported in this study are publicly available at \url{https://github.com/IvanGilardoni/Jeffreys_prior}.

\bibliographystyle{aipnum4-2}
%
\providecommand{\noopsort}[1]{}\providecommand{\singleletter}[1]{#1}%

\clearpage

\renewcommand{\thefigure}{S\arabic{figure}}
\renewcommand{\thepage}{S\arabic{page}} 

\setcounter{figure}{0}
\setcounter{page}{1}

\onecolumngrid

\section*{Supplementary Material}

\subsection{Implementation details}

The methodology described in this work has been implemented in the Python package \texttt{MDRefine} through two functions:
\texttt{local\_density} and \texttt{posterior\_sampling}.
The function \texttt{local\_density} computes the local density of ensembles in the cases of ensemble refinement or force-field fitting. It requires three input arguments: \texttt{variab}, containing the quantities needed to compute the density (such as observables or generalized forces, described in the following); \texttt{weights}, containing the normalized weights corresponding to the probability distribution at which the density is evaluated; and \texttt{which\_measure}, which specifies the prescription for the local density.
By default, the ensemble density is computed using the Jeffreys measure (\texttt{which\_measure='jeffreys'}). Alternative choices include the BELT2 measure (\texttt{which\_measure='average'}) and the Dirichlet measure (\texttt{which\_measure='dirichlet'}).
The function returns the local ensemble density together
with the matrix entering the density computation (the covariance matrix for Jeffreys and BELT2).

In the default case \texttt{which\_measure = `jeffreys'}, the local density is $\sqrt{\det C}$ with $C$ the variance-covariance matrix of either the observables $C(\lambda)$ in ensemble refinement or the generalized forces $C(\phi)$ in force-field fitting (see Eqs. \ref{eqn:der_Fisher}--\ref{eqn:dn}).
In this case, the argument \texttt{variab} is either the array of observables \texttt{g} (ensemble refinement) or the tuple \texttt{(fun\_forces, pars, f)} (force-field fitting), where \texttt{fun\_forces} computes the generalized forces $\partial V/\partial \phi_j$, \texttt{pars} contains the parameters $\phi$, and \texttt{f} contains the quantities required to evaluate the forces.

\begin{lstlisting}[language=Python]
from MDRefine.bayesian import local_density

# Ensemble refinement
measure, cov = local_density(g, weights)

# Force-field fitting
V = ff_correction(pars, f)

fun_forces = jax.jacfwd(ff_correction, argnums=0)
my_tuple = (fun_forces, pars, f)
measure, cov = local_density(my_tuple, weights)
\end{lstlisting}

For \texttt{which\_measure='average'} (BELT2), the local density is computed from the determinant of the covariance matrix. In ensemble refinement, this corresponds to the covariance matrix of the observables, whereas in force-field fitting, it corresponds to the covariance matrix of observables and generalized forces. Accordingly, \texttt{variab} is either the observable array \texttt{g} or the tuple \texttt{(fun\_forces, pars, f, g)}.

Finally, in the case \texttt{which\_measure = `dirichlet'} the input \texttt{variab} is the same as in the first case \texttt{which\_measure = `jeffreys'}. 

Since the matrices involved are real, symmetric, and positive semidefinite, determinants are computed through a Cholesky decomposition, which is numerically efficient for large matrices. Writing
\[
M = LL^{T},
\]
where \(L\) is a lower triangular matrix, one obtains
\[
\sqrt{\det M} = \det L = \prod_i L_{ii}.
\]
In the current version of the code, if the Cholesky decomposition fails because the numerically computed matrix is not positive definite, the matrix is symmetrized and shifted by $(-\lambda_{\min}+10^{-8})I$, where $\lambda_{\min}$ is its smallest eigenvalue, before repeating the decomposition. This correction is not expected to be invoked under ordinary sampling conditions, unless the selected observables are numerically linearly dependent or the multipliers become so large that the effective ensemble collapses onto too few configurations and the covariance loses numerical rank.

These functions are implemented in the
\href{https://github.com/bussilab/MDRefine/blob/master/MDRefine/bayesian.py}{\texttt{bayesian.py}}
module of \texttt{MDRefine}. The module also contains the function \texttt{posterior\_sampling}, which performs Metropolis Monte Carlo sampling of the posterior distribution for both ensemble refinement and force-field fitting. Usage examples are provided in the Jupyter notebook 
\href{https://github.com/bussilab/MDRefine/blob/master/Examples/Tutorial_sampling.ipynb}{\texttt{Tutorial\_sampling.ipynb}}.

We benchmarked the dependence of the sampling cost on the number of restrained observables using the RNA calculation at stride 100 (10,170 retained MD frames). Timings were obtained on one CPU core of the Leonardo supercomputer from 2,000 MCMC steps, after one untimed initialization call, and are reported as the median of three independent runs. For flat BELT, the time increased from 5.18~ms per step with one observable to 11.95~ms with 28 observables. For Jeffreys, it increased from 5.38 to 25.89~ms per step over the same range. Thus, with 28 observables the covariance-matrix and determinant calculation increases the cost by a factor of approximately 2.2 relative to flat BELT. BELT2 has effectively the same cost as Jeffreys because both require the same covariance matrix and determinant and differ only in the power applied to the determinant.

\clearpage
\subsection{Supplementary results for the Gaussian model}

\begin{figure}
\centering
\subfloat[]{\includegraphics[width=0.8\textwidth]{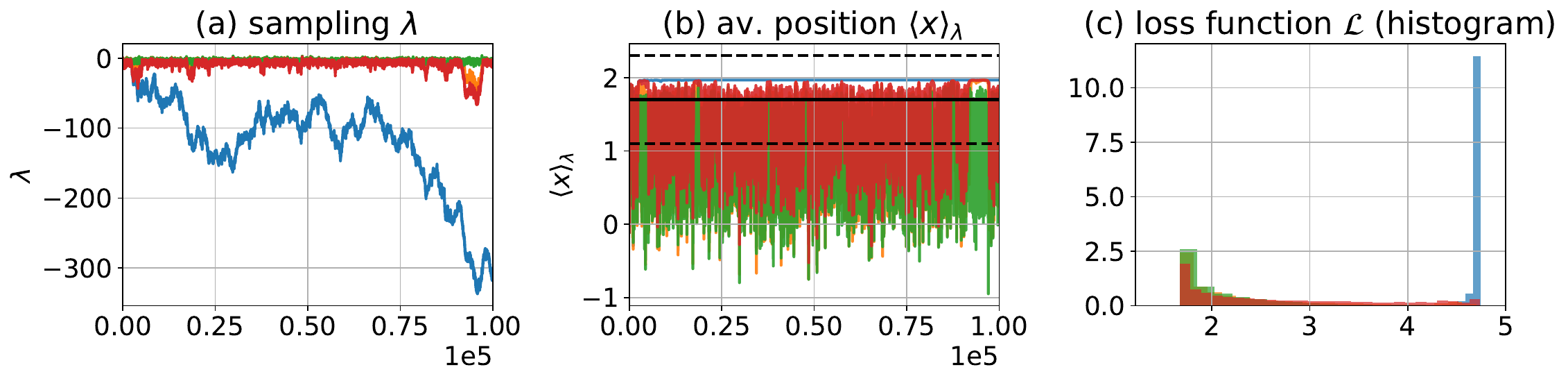}}
\hfill
\subfloat[]{\includegraphics[width=0.8\textwidth]{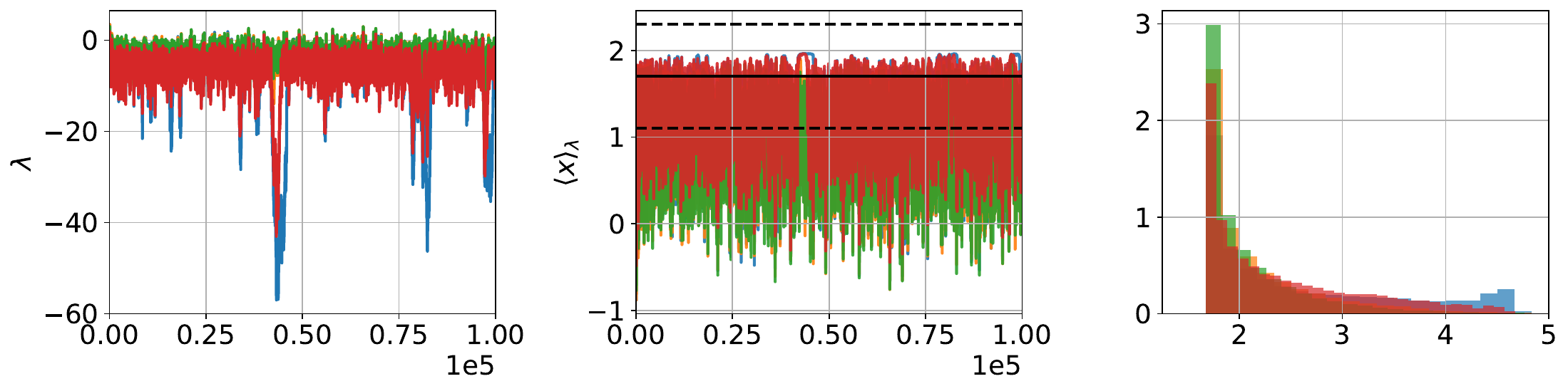}}
\hfill
\subfloat[]{\includegraphics[width=0.8\textwidth]{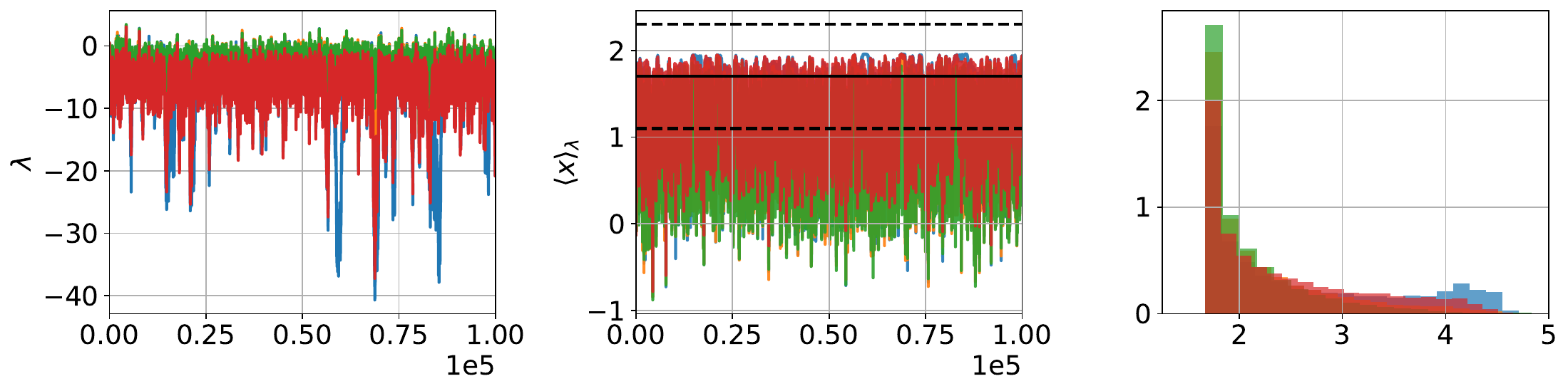}}
\hfill
\subfloat[]{\includegraphics[width=0.8\textwidth]{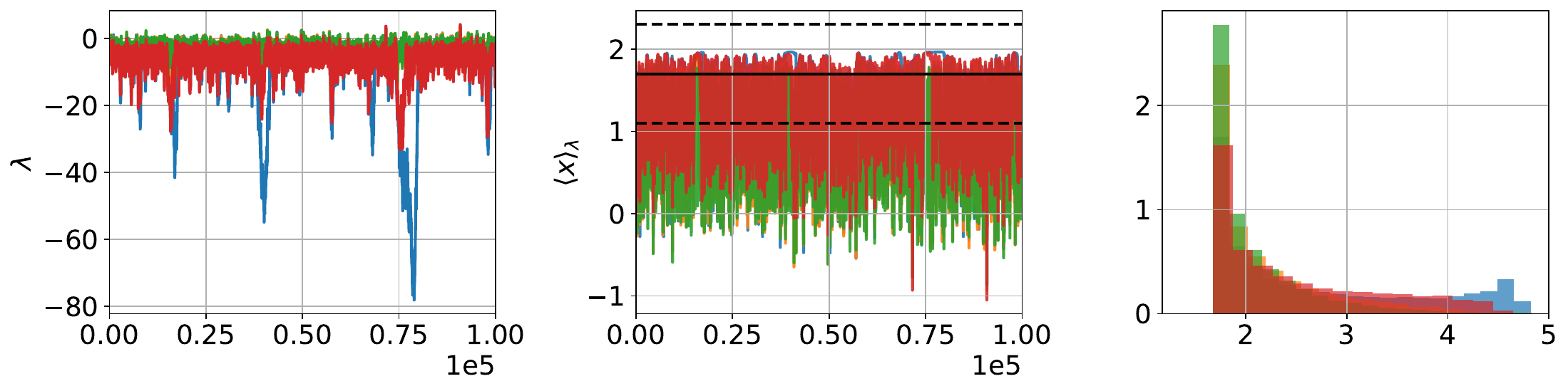}}
\hfill
\subfloat[]{\includegraphics[width=0.8\textwidth]{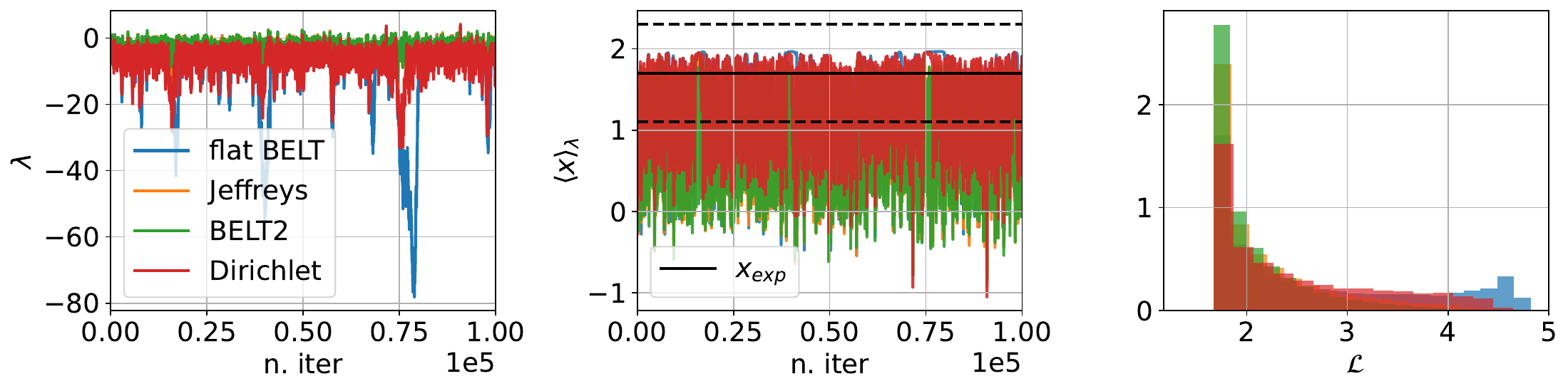}}
\caption{Additional Monte Carlo samplings of the Gaussian model using different random seeds. Panels correspond to Fig.~\ref{fig:example1_sampling}(a–c) of the main text. The variability among trajectories illustrates the stochastic nature of the excursions into the plateau region when the flat BELT measure is used, whereas the Jeffreys, BELT2, and Dirichlet measures remain localized near the region of largest posterior mass.
\label{fig:ex1_different_seeds}
}
\end{figure}

\begin{figure}
    \centering
    \includegraphics[width=1\linewidth]{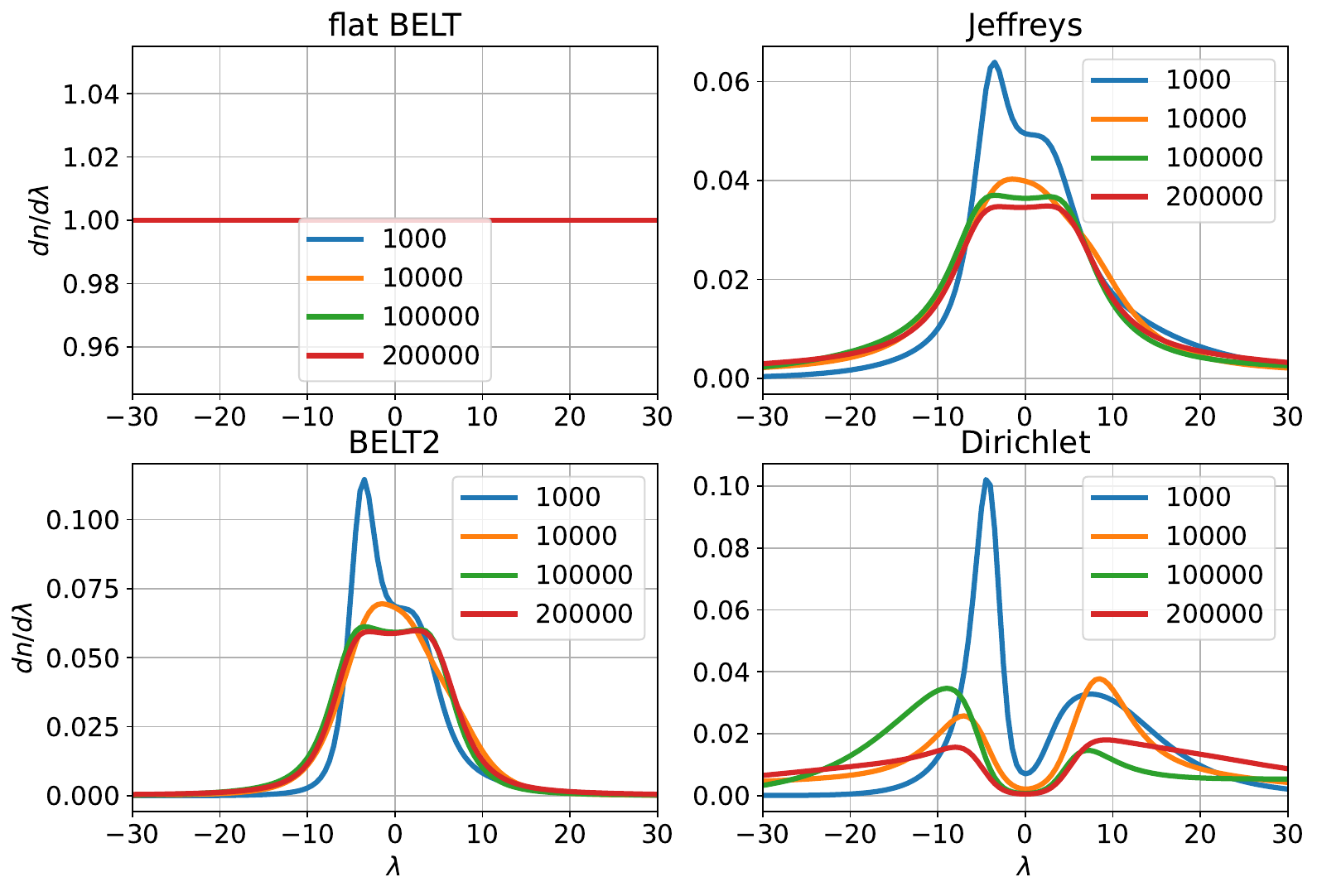}
    \caption{Ensemble-counting densities for the Gaussian model obtained with different numbers of sampled frames. The flat BELT measure remains uniform in $\lambda$ by construction. The Jeffreys and BELT2 densities become increasingly localized as the number of frames increases, reflecting the growing separation between distinguishable ensembles. The Dirichlet density develops a pronounced minimum near $\lambda \approx 0$ for large numbers of frames because of the discrete representation of the reference ensemble.}
    \label{fig:several_strides}
\end{figure}

We here report supplementary results for the Gaussian model discussed in Section \ref{sec:quadratic}. Figure~\ref{fig:ex1_different_seeds} shows additional Monte Carlo trajectories obtained with different random seeds, corresponding to the analysis in Fig. \ref{fig:example1_sampling}(a–c) of the main text. Figure~\ref{fig:several_strides} shows how the ensemble-counting densities depend on the number of sampled frames, extending the comparison presented in Fig.~\ref{fig:example1_sampling}(d).
\clearpage

\clearpage
\subsection*{Supplementary figures for the RNA oligomer}

\begin{figure}
\includegraphics[width=1\linewidth]{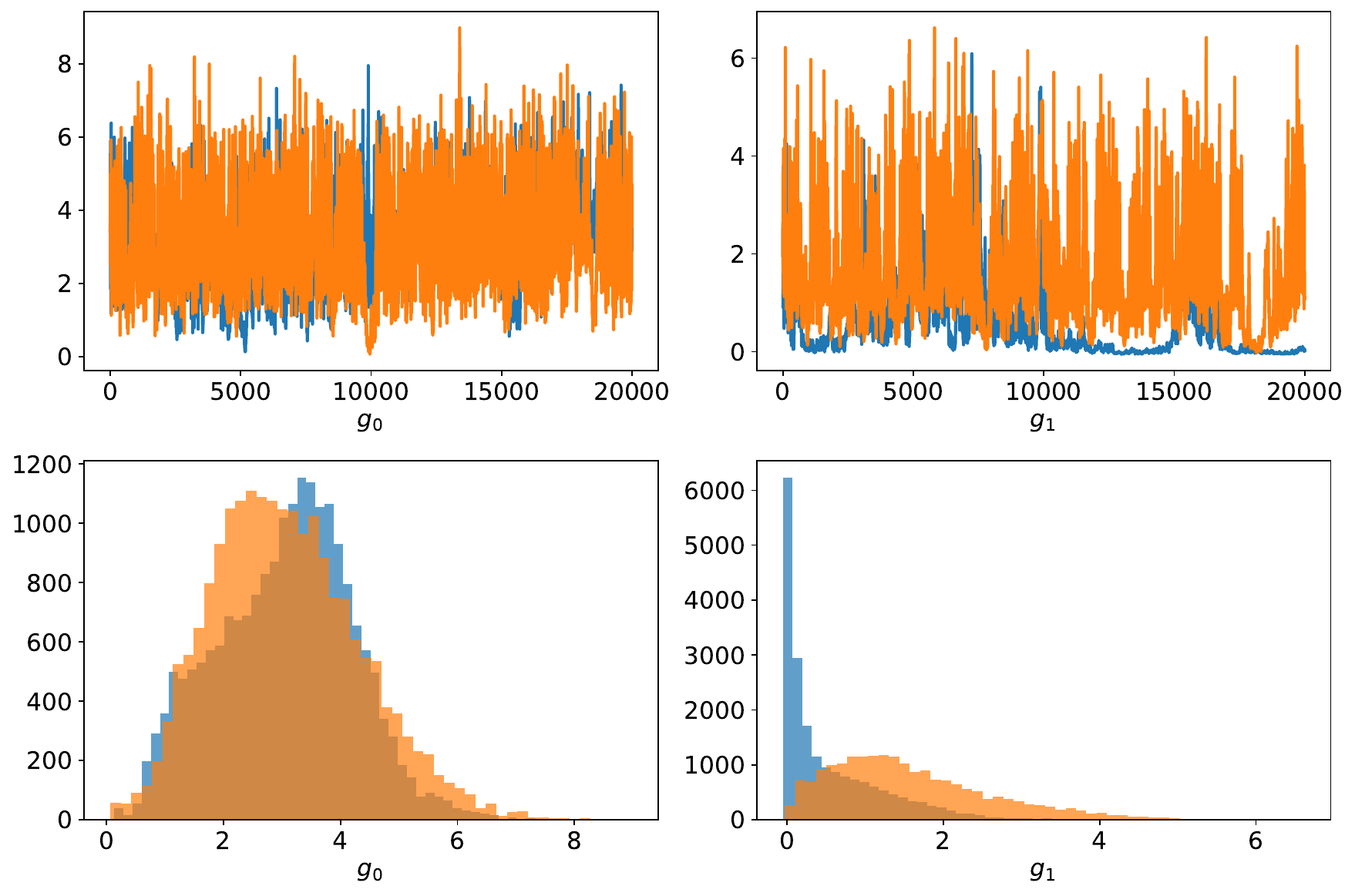}
\caption{
Time series and marginal histograms of the two restrained observable averages sampled in the two-dimensional RNA refinement shown in Fig.~\ref{fig:example2}. The flat BELT and Jeffreys measures lead to different posterior distributions in observable space, consistent with the broader plateau exploration observed for the flat measure in Fig.~\ref{fig:example2}.}
\label{fig:example2_si}
\end{figure}

\begin{figure}
\centering
\subfloat[]{\includegraphics[width=0.9\textwidth]{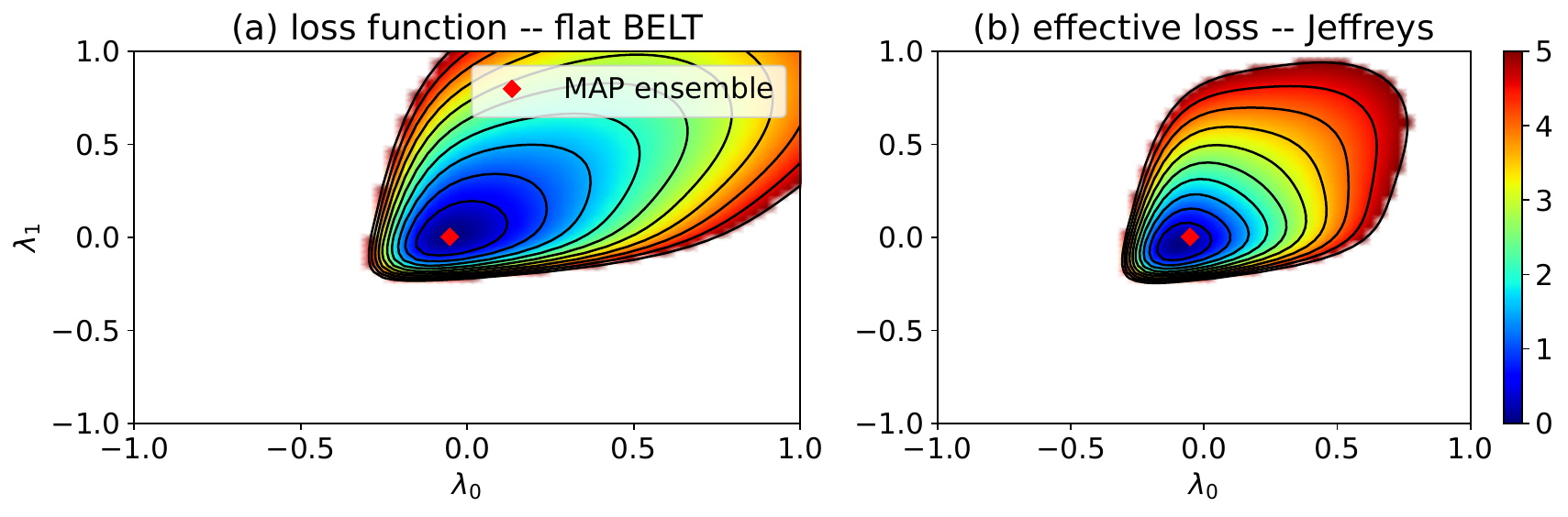}}
\hfill
\subfloat[]{\includegraphics[width=0.9\textwidth]{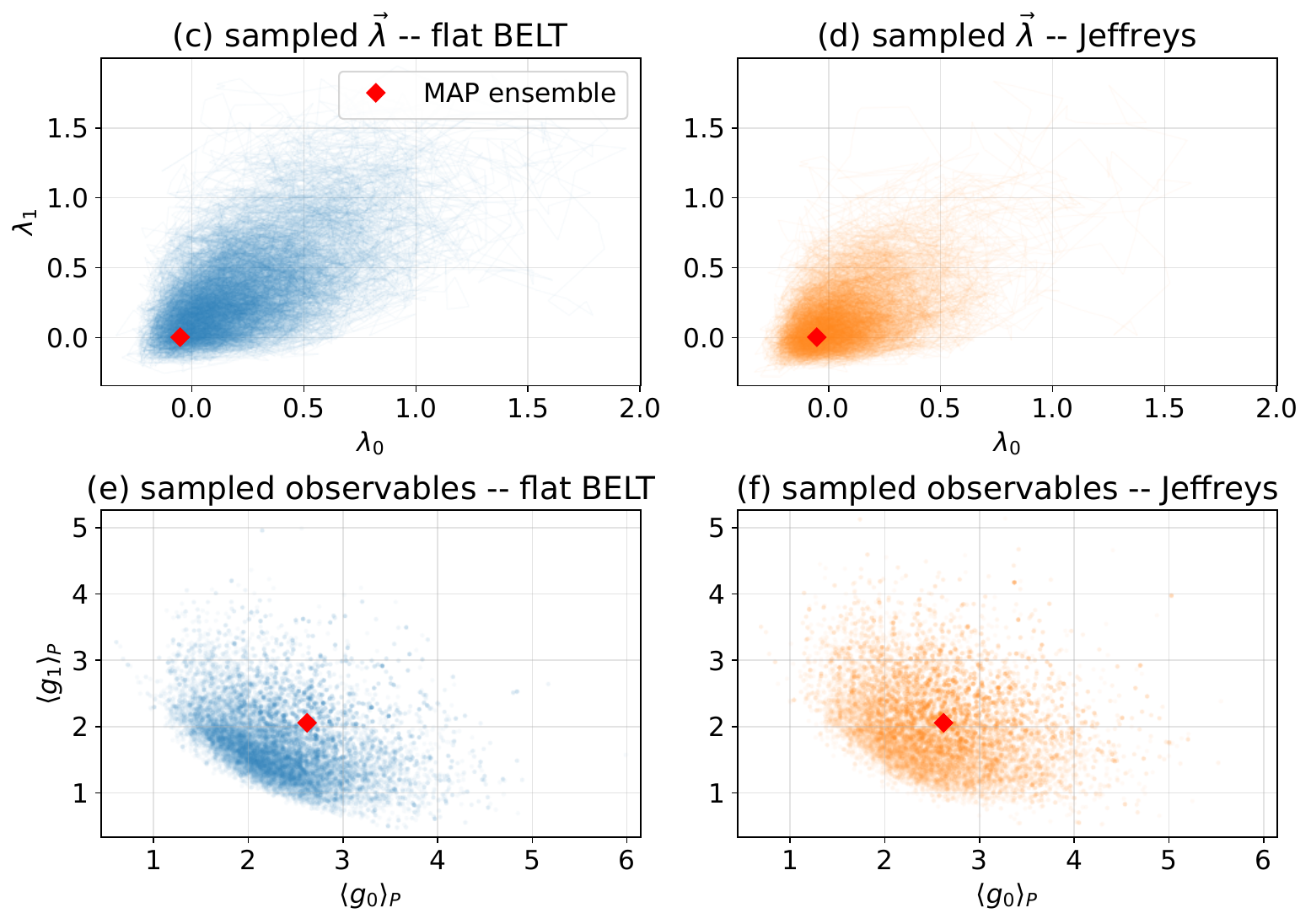}}
\caption{Same analysis as Fig.~\ref{fig:example2} of the main text, but for
a larger regularization hyperparameter ($\alpha=10$; stride = 10, corresponding to 101,700 MD frames). The stronger
regularization increases the loss difference between the posterior
minimum and the plateau region, making excursions into the plateau
less likely during sampling.}
\label{fig:example2_alpha10}
\end{figure}

\begin{figure}
\centering
\subfloat[]{\includegraphics[width=1\linewidth]{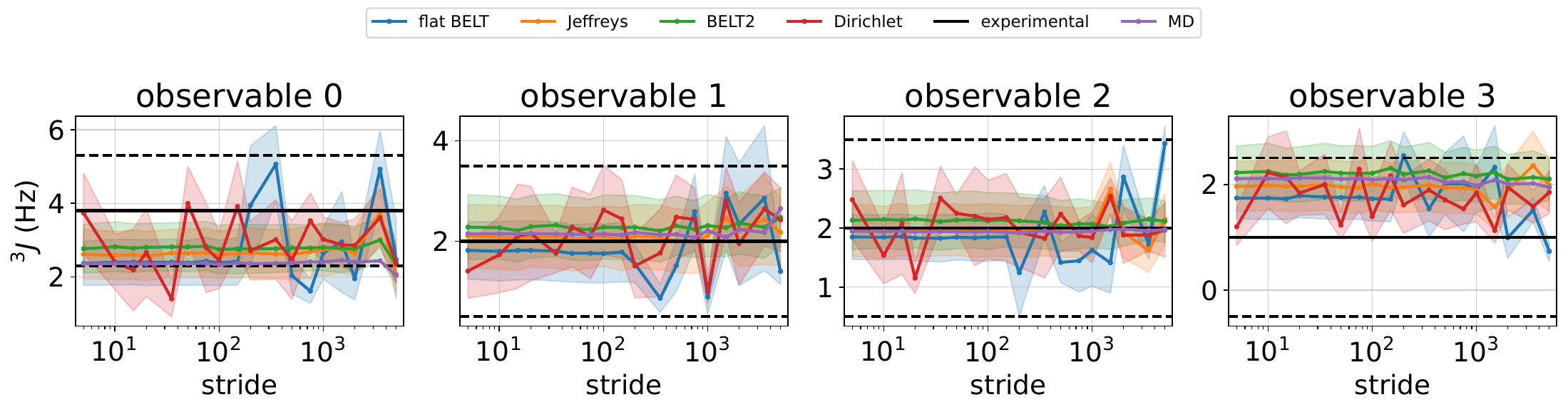}}
\hfill
\subfloat[]{\includegraphics[width=1\linewidth]{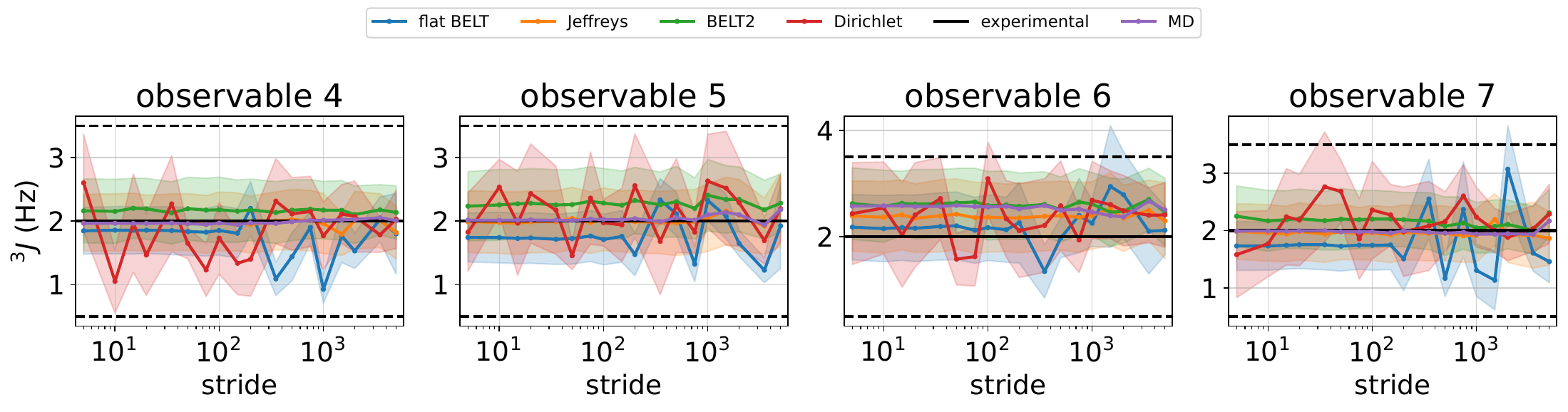}}
\hfill
\subfloat[]{\includegraphics[width=1\linewidth]{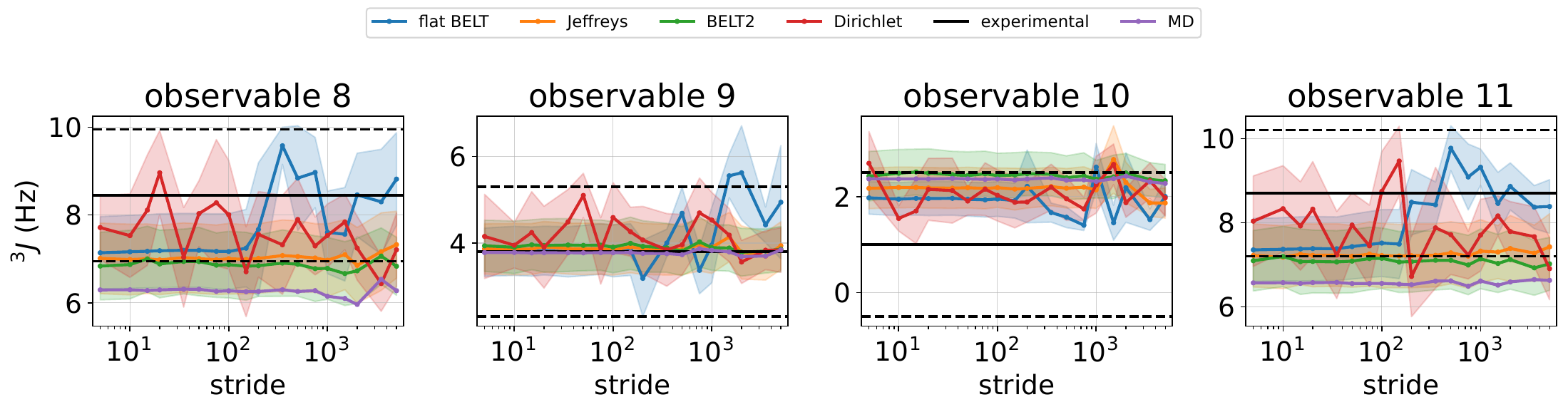}}
\hfill
\subfloat[]{\includegraphics[width=1\linewidth]{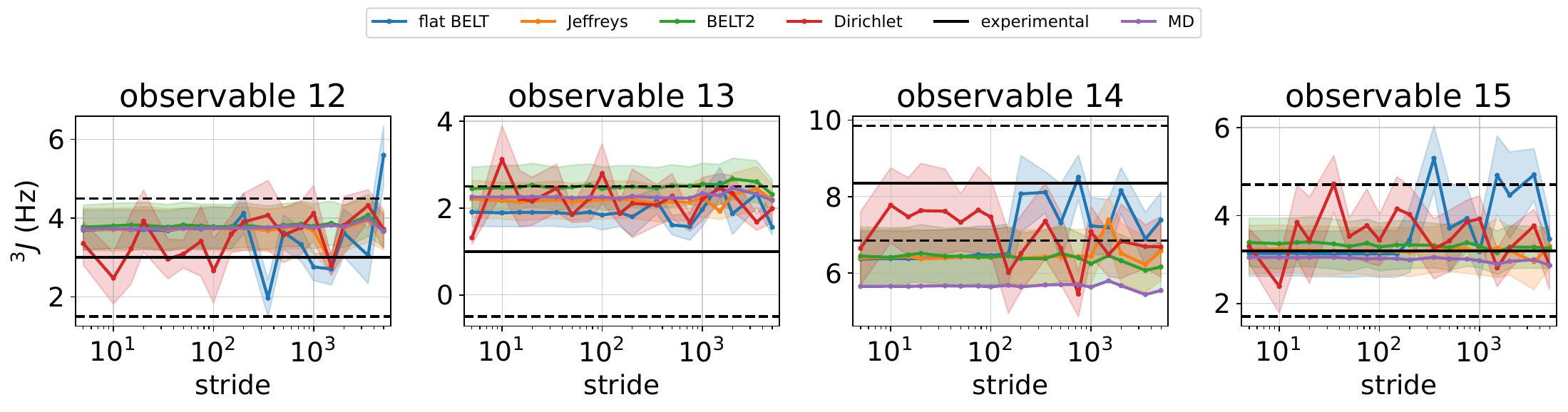}}
\caption{Same analysis as in Fig.~\ref{fig:example2b}a, for observables 0--15.}
\label{fig:other_obs1}
\end{figure}

\begin{figure}
\centering
\subfloat[]{\includegraphics[width=1\linewidth]{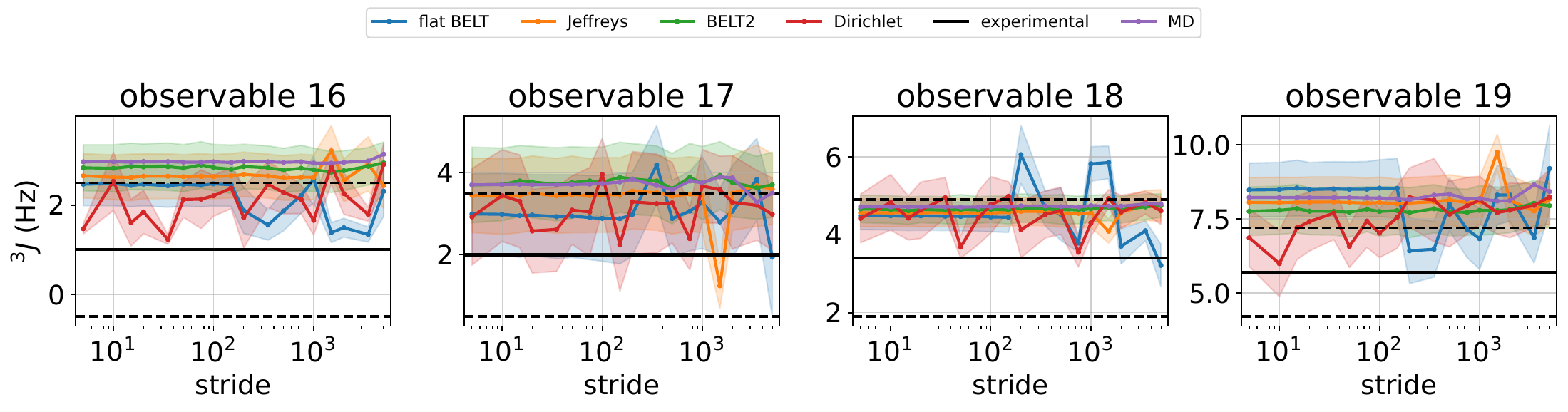}}
\hfill
\subfloat[]{\includegraphics[width=1\linewidth]{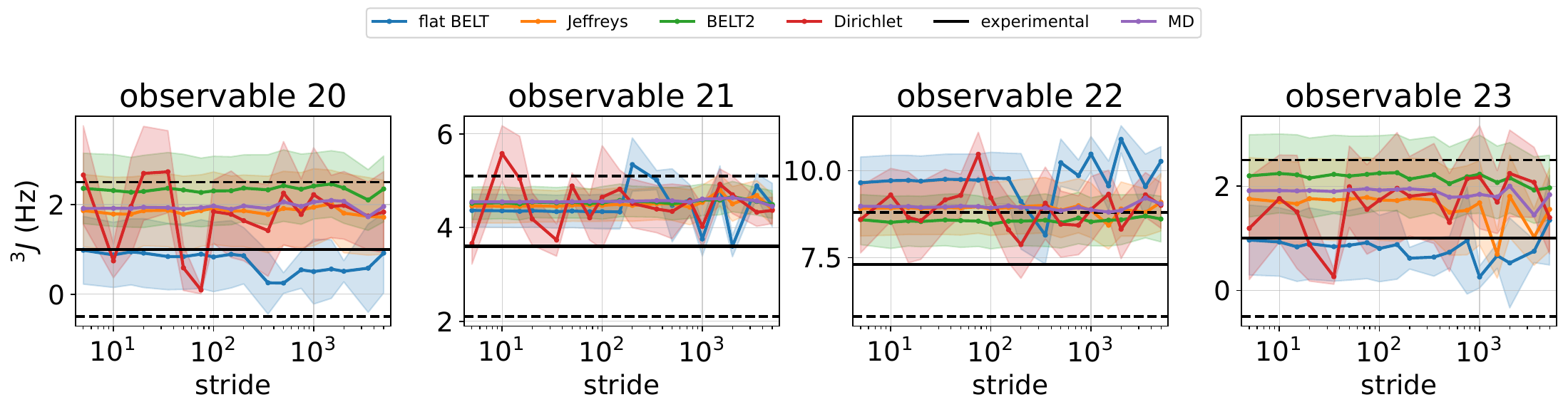}}
\hfill
\subfloat[]{\includegraphics[width=1\linewidth]{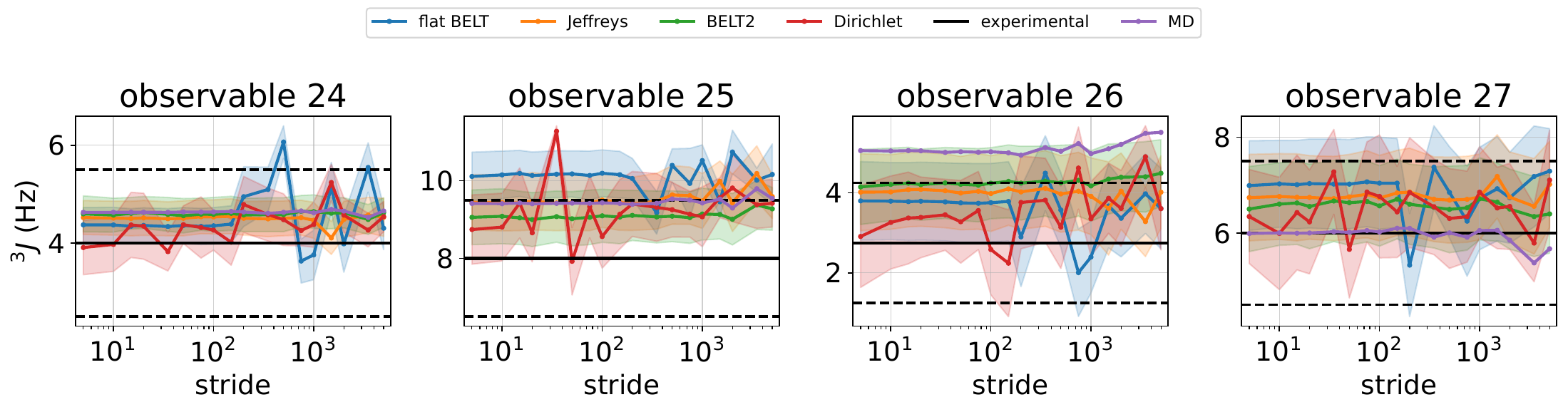}}
\caption{Same analysis as in Fig.~\ref{fig:example2b}a, for observables 16--27.}
\label{fig:other_obs2}
\end{figure}

We here report supplementary results for the RNA oligomer test case discussed in Section \ref{sec:rna-oligomer}. Figure~\ref{fig:example2_si} shows the time series and marginal distributions of the restrained observables sampled in the two-dimensional posterior shown in Fig.~\ref{fig:example2}. Figure~\ref{fig:example2_alpha10} provides additional two-dimensional posterior visualizations for a larger regularization parameter ($\alpha=10$).
Figures~\ref{fig:other_obs1} and~\ref{fig:other_obs2} report posterior averages for all 28 scalar couplings as a function of trajectory stride, extending the representative observable shown in Fig.~\ref{fig:example2b}a.

\clearpage

\begin{figure}
\centering
\includegraphics[width=0.9\linewidth]{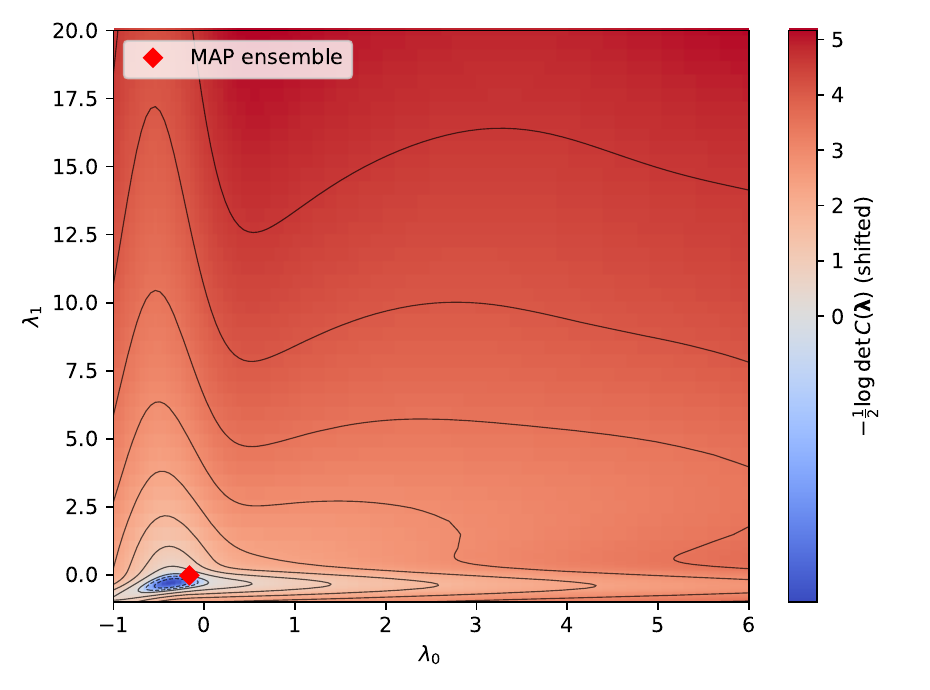}
\caption{Difference between the effective losses obtained with the Jeffreys and flat BELT ensemble-counting measures for the two-observable RNA example of Fig.~\ref{fig:example2}. The plotted quantity is $\mathcal L_{\mathrm{Jeffreys}}-\mathcal L_{\mathrm{flat}}=-\frac{1}{2}\log\det C(\vec\lambda)$, up to an additive constant chosen so that it is zero at the MAP ensemble (red diamond). The covariance matrix was evaluated from the same finite reference ensemble used in Fig.~\ref{fig:example2} (stride 10; 101,700 retained MD frames), over the same $\vec\lambda$ domain. The correction increases toward the collapsed-ensemble plateau.}
\label{fig:example2_correction}
\end{figure}

\end{document}